\documentclass[twocolumn,trackchanges]{aastex701}

\usepackage{multirow}
\usepackage{amsmath}  
\usepackage{graphicx}  
\usepackage{enumitem}  

\begin{document}

\title{Dual AGNs on 100 kpc Scales from the Million Quasar Catalog}

\author[orcid=0009-0008-8080-3124]{Zhuojun Deng}
\altaffiliation{Co-author with equal contribution.}
\email{202431101076@mail.bnu.edu.cn}
\affiliation{School of Physics and Astronomy, Beijing Normal University, Beijing 100875, China}
\affiliation{Institute for Frontier in Astronomy and Astrophysics, Beijing Normal University, Beijing, 102206, China}

\author[orcid=0009-0000-6610-8979]{Cheng Xiang}
\altaffiliation{Co-author with equal contribution.}
\email{202221160024@mail.bnu.edu.cn}
\affiliation{School of Physics and Astronomy, Beijing Normal University, Beijing 100875, China}
\affiliation{Institute for Frontier in Astronomy and Astrophysics, Beijing Normal University, Beijing, 102206, China}

\author[orcid=0009-0006-9345-9639]{Qihang Chen}
\email{202131160006@mail.bnu.edu.cn}
\affiliation{School of Physics and Astronomy, Beijing Normal University, Beijing 100875, China}
\affiliation{Institute for Frontier in Astronomy and Astrophysics, Beijing Normal University, Beijing, 102206, China}

\author[orcid=0000-0003-1188-9573]{Liang Jing}
\email{202331160009@mail.bnu.edu.cn}
\affiliation{School of Physics and Astronomy, Beijing Normal University, Beijing 100875, China}
\affiliation{Institute for Frontier in Astronomy and Astrophysics, Beijing Normal University, Beijing, 102206, China}

\author[orcid=0009-0008-9072-4024]{Xingyu Zhu}
\email{202321160028@mail.bnu.edu.cn}
\affiliation{School of Physics and Astronomy, Beijing Normal University, Beijing 100875, China}
\affiliation{Institute for Frontier in Astronomy and Astrophysics, Beijing Normal University, Beijing, 102206, China}

\author{Jianghua Wu$^\dagger$}
\email[show]{jhwu@bnu.edu.cn}
\email{Corresponding author.}
\affiliation{School of Physics and Astronomy, Beijing Normal University, Beijing 100875, China}
\affiliation{Institute for Frontier in Astronomy and Astrophysics, Beijing Normal University, Beijing, 102206, China}

\begin{abstract}
    Research on dual active galactic nuclei (AGNs) is crucial for understanding the coevolution of galaxies and supermassive black holes. However, the current number of dual AGNs remains scarce.
    In this work, we selected 173 new dual AGNs, 4 AGN triplets, and 1 AGN quadruplet from the Million Quasars Catalog, all with low redshift ($z < 0.5$), a projected distance ($r_p$) of no more than 100 kpc, and a line-of-sight velocity difference ($|\Delta v|$) of less than 600 km s$^{-1}$, thus supplementing existing low-redshift dual AGNs demographics. Visual inspection of the optical images from the Dark Energy Spectroscopic Instrument Legacy Survey was performed for each pair, revealing that $\sim$16\% of pairs exhibit tidal features.
    Statistical analyses show an increasing number of dual AGNs with decreasing redshift, with velocity difference primarily at $|\Delta v| < $ 300 km s$^{-1}$, which is likely an artifact of our selection strategy. The tidal sample peaks as having 13 pairs at 5-20$h^{-1}_{70}$ kpc, but drops to 1 pair $> 55\,h^{-1}_{70}$ kpc. Our study also explores thewide separation ($r_p>10$ kpc) dual AGNs, finding 165 such systems, with 25 displaying clear tidal features. Furthermore,
    some extra galaxies, AGNs, and/or their candidates were found in the same regions of the pairs or
    multiplets forming interacting systems with these pairs or multiplets.
\end{abstract}

\keywords{\uat{Black hole physics}{1879} --- \uat{Galaxies}{573} --- \uat{Interactions galaxies}{600}  --- \uat{Active galactic nuclei}{16}}

\section{Introduction}
    Galaxy interactions and mergers have long been recognized as critical factors in shaping the structure and morphology of galaxies \citep{Zwicky1956, Toomre&Toomre1972, Sanders1988, Barnes&Hernquist1992, Lotz2008, Conselice2014}. They usually manifest themselves as dual or even multiple cores, irregular galactic morphologies, gas bridges or rings, and tidal tails. These interactions cause the galactic gas to lose angular momentum and to move towards the galactic center \citep{Barnes1996, Cox2008, Blumenthal2018}.
    If the gas can reach the supermassive black holes (SMBHs), they can begin fueling as active galactic nuclei (AGNs) \citep{Kormendy2013, Hopkins2010}; otherwise, the gas may instead trigger starburst activity \citep{DiMatteo2005Nature, Hopkins2008, Cox2008}.

    The study of dual AGNs in galaxy mergers is of great importance to various astrophysical inquiries. They provide unique laboratories for the investigation of galaxy evolution, star formation, the growth and evolution of SMBHs, and the eventual emission of gravitational waves \citep{Hopkins2006ApJS, Kauffmann2000, DiMatteo2005Nature}.
    SMBH pairs are thought to be a generic outcome in the hierarchical paradigm of structure formation \citep{Begelmanetal1980, Merritt2001ApJ, Yu2002MNRAS}, given that most massive galaxies harbor SMBHs \citep{Kormendy1995, Richstoneetal1998}.
    Both SMBHs in a merger may accrete at the same time, which will be observable as a pair of AGNs. Those pairs with separations of individual members $<100$ kpc are also frequently referred to as "dual AGNs" \citep{DeRosaetal2019, Pfeifle2025}.      
    However, it is difficult to predict when one, and especially both, SMBHs become active \citep{Shlosmanetal1990, Armitage2002, Wada2004, Dottietal2007, Dottietal2012}. 
    The frequency of dual AGNs can constrain models involving galaxy merger rates and tidally triggered AGN activity \citep{Yuetal2011, Foord2024}. The properties of their host galaxies offer clues about the external and internal conditions that enable both SMBHs to become active \citep{Hopkins2008, Blecha2013}.
    When two galaxies merge, their SMBHs sink via dynamical friction, form a bound binary, harden, and finally coalesce \citep{Begelmanetal1980}. The merger of binary SMBHs is anticipated to produce low-frequency gravitational waves \citep{Reardon2023PPTA, 2024EPTA_Collaboration}, the detection of which would directly test the theory of general relativity and allow for the exploration of SMBH populations in the early universe \citep{Hawking1987, Haehnelt1994, Holz2005, Arzoumanian2020ApJ, LISA2023}.

    There have been an increasing number of studies that identified dual AGNs with separations ranging from $<$ 1 kpc to tens of kpc in the X-ray \citep{Komossaetal2003, Balloetal2004, Bianchietal2008, Piconcellietal2010, Mazzarella2011, Koss2012ApJL, Kossetal2016}, radio \citep{Owenetal1985, Rodriguez2006, Fu2011b}, and optical \citep{Myers2006ApJ, Myers2007ApJ, Comerfordetal2009b, Hennawi2006, Hennawi2010, Liuetal2010a, Liu2011ApJ} bands. Moreover, several tens of quasar pairs have also been discovered as byproducts of gravitational lensing searches \citep{Inada2012, Lemon2018, Yue2023, Dux2024}. 
    Due to the extensive sky coverage, wide wavelength range, high data quality, and rich sample of Sloan Digital Sky Survey (SDSS) spectra, the SDSS is commonly used to systematically identify and analyze dual AGN candidates.
    A subset of systems on the kpc scale identified in SDSS spectra has been further confirmed through high-resolution near-infrared(NIR) imaging \citep{Rosarioetal2011, DingXH2025} and spatially resolved optical spectroscopy \citep{Liuetal2010a, Fuetal2012}.
    
    These objects were initially identified through the “double-peak” selection method \citep{Wangetal2009, Liuetal2010b, Shietal2014, Wangetal2017}, which searches for double-peaked emission-line profiles as indicators of dual nuclei. However, this method exhibits two limitations: firstly, due to the limited resolution of SDSS spectra, it is biased against pairs with line-of-sight (LOS) velocity differences smaller than approximately 150 km s$^{-1}$ \citep{Liuetal2010a, Liuetal2010b}. Secondly, it is biased against pairs with separations larger than a few kpc, as pairs with angular distances greater than $3^{\prime\prime}$ (the diameter of an SDSS fiber) cannot fit within a single SDSS fiber. 

    To circumvent the limitations of the double-peak method and establish a larger sample of dual AGNs at low redshift, \citet{Liu2011ApJ} adopted complementary methods to identify AGN pairs at $\bar{z} \sim 0.1$ from SDSS DR7. They assembled a sample of 1286 dual AGNs and candidates, 30\% of which show morphological tidal features in SDSS optical images.
    In recent years, the number of AGNs and quasars has significantly increased. \citet{Pfeifle2025} presented the Big Multi-AGN Catalog (The Big MAC) DR1, which includes 156 confirmed dual AGNs and 4180 dual AGN candidates, compiled from hundreds of literature articles published between 1970 and 2020. As of June 2023, the Million Quasar Catalog (MQC, version 8.0; \citealt{Flesch2023OJA}) has cataloged 1,021,800 quasars and candidates. This provides new possibilities for further exploration of low-redshift dual AGNs.

    We conducted a systematic search for multiple AGN systems within the range of $0 < z < 0.5$ in the MQC, limiting the projected distance to $r_p < 100h^{-1}_{70}$ kpc and the LOS velocity difference to $|\Delta v| < 600$ km s$^{-1}$. The latter two selection criteria were proposed by \citet{Silvermanetal.2011, Liu2011ApJ}. These criteria are also consistent with prior works and definitions for dual AGNs \citep{Koss2011ApJ, Koss2012ApJL, DeRosaetal2019}, and this is similar to the dual AGN criteria recently laid out by \citet{Pfeifle2025}. The objective is to identify new multiple AGN systems at low redshifts, and to study the characteristics of dual AGNs in order to understand the process of galaxy mergers and their impact on galaxy evolution.

    This paper is organized as follows: In Section \ref{sec2}, the data and the method of sample selection are described. We present images of dual AGNs from the Ninth or Tenth Data Release of the Dark Energy Spectroscopic Instrument Legacy Survey\footnote{\url{https://www.legacysurvey.org/acknowledgment/}} (DESI–LS; \citet{DESI-LS2019}). In Section \ref{sec3}, we examine the frequency of dual AGNs with tidal features and provide preliminary statistics of these pairs. Section \ref{sec4} and \ref{sec5} give the discussion on the implications of our results and the conclusions. Throughout, a $\Lambda$-cold dark matter ($\Lambda$CDM) cosmology is assumed with $\Omega_{m} = 0.3$, $\Omega_{\Lambda} = 0.7$, and H$_{0} = 70 \ \text{km s}^{-1}\text{Mpc}^{-1}$.

\section{Data And Sample Selection} \label{sec2}
\subsection{Data} \label{subsec21}
    All of our objects come from the MQC, Version 8, which contains 907,144 type-I QSOs and AGNs. 66,026 QSO candidates are also included, calculated via radio/X-ray association (including double radio lobes) with a confidence level of 99\%. Including blazars and type II objects increases the total number of AGNs to 1,021,800. Compared to previous studies of dual AGNs \citep{Liu2011ApJ, Koss2012ApJL}, our parent sample is larger by several orders of magnitude, but it is intrinsically heterogeneous and drawn from a variety of surveys. Therefore, it is important to give a detailed description of its classification scheme and redshift assignments, which are directly relevant to the construction of our sample.
    
    In the MQC, sources classified as Q (QSO, broad-line nucleus–dominated quasars), A (AGNs, broad-line Seyferts/host galaxy–dominated AGNs), K (narrow-line QSO), and N (narrow-line Seyfert) are all assigned spectroscopic redshifts. 
    For B-class objects (BL Lac type), redshift information is not required and may therefore be absent, or represented by a photometric redshift with a typical precision of $\sim$0.1.
    More than 90\% of the AGNs in our sample consist of broad-line and narrow-line Seyfert populations, which are mainly drawn from the PGC\footnote{\url{https://heasarc.gsfc.nasa.gov/W3Browse/all/pgc2003.html}} and SDSS DR16 catalogs.
    We emphasize that, given the MQC inclusion rules, it is appropriate to describe our working sample drawn from MQC as spectroscopically confirmed: when literature samples mix spectroscopic and photometric redshifts, MQC retains only spectroscopic quasars as classified objects to ensure reliable classifications and redshifts \citep{Flesch2023OJA}. Accordingly, all sources in our sample have spectroscopic redshifts. Note that MQC uses a broad “AGN” label and does not distinguish Seyferts, LINERs, and Composites. Thus, we assume LINERs and Composites represent AGNs, consistent with \citet{Liu2011ApJ}. Although some LINERs/Composites may in fact not host actively accreting SMBHs, we assume that these systems are accreting SMBHs in this work.

    In addition, the suffixes R (radio), X (X-ray), and 2 (double radio lobes) indicate that the source has been associated with detections in the corresponding bands, based on the Millions of Optical-Radio/X-ray Associations (MORX) catalogue \citep{Flesch2024MORX}. The latest MORX v2 release evaluates associations from all the largest radio and X-ray surveys to June 2023, including VLASS, LoTSS, RACS, FIRST, NVSS, and SUMSS in radio, and Chandra, XMM-Newton, Swift, and ROSAT in X-rays \citep{Flesch2024MORX}. 
    These associations are based on optical–radio/X-ray cross-matching probabilities, derived by comparing the areal densities of optical profiles around radio/X-ray detections to background averages \citep{Flesch&Hardcastle2004}. 
    It should be noted that composite labels in the MQC (e.g., AR, NX, NR2) primarily indicate the availability of multiwavelength observations (radio/X-ray/double radio lobes) and do not alter the intrinsic physical classification of the source, nor do they imply that the objects are bona fide radio/X-ray AGNs.

    For consistency and to simplify the statistical analysis, we adopt the following classification scheme: 
    (1) sources labeled with "Q" (e.g., Q, QR, QRX, QX) are uniformly classified as QSOs; 
    (2) sources labeled with "A" (e.g., A, AR, AX, ARX, AR2, AR2X) are grouped into the AGN category (broad-line Seyfert-type AGNs); 
    (3) sources whose class string begins with “B” (e.g., B, BR, BX, BRX, B2) are classified as BL Lac objects (FSRQs are instead typed as QSOs);
    (4) sources labeled with "K" or "KX" are grouped into the NLQSO class (type-II narrow-line QSOs); 
    (5) sources labeled with "N", "NR", "NRX", "NX", or "NR2" are grouped into the NLAGN class (narrow-line Seyfert-type or host-dominated AGNs). The majority of these narrow-line objects in our sample originate from SDSS DR16, but a non-negligible fraction are drawn from other spectroscopic surveys such as 2dF, 2SLAQ, and the Principal Galaxies Catalogue (PGC).
    
    Our optical imaging data are from DESI-LS, a project dedicated to constructing a comprehensive map of the sky through optical and infrared imaging observations. 
    These images were used in the visual examination of our sample in the optical band, specifically in identifying the signs of galaxy interactions. In comparison to SDSS and the Panoramic Survey Telescope and Rapid Response System (Pan-STARRS), DESI-LS stands out for its extensive depth and coverage, making it superior for tasks such as galaxy morphological classification.

\subsection{Sample Selection} 
\label{subsec22}

\noindent
    Our sample of dual AGNs was selected based on three criteria: (1) The redshift cut of 0 $<$ \textit{z} $<$ 0.5, (2) The projected distance between individual members of $r_p < 100\,h^{-1}_{70}$ kpc, and (3) the line of sight (LOS) velocity difference between members of $|\Delta v| < 600\,\mathrm{km\, s^{-1}}$ \citep{Silvermanetal.2011, Liu2011ApJ}. The choice of $100\,h^{-1}_{70}$ kpc is somewhat larger than the typical thresholds adopted in galaxy pair studies \citep{Bartonetal.2000, Ellisonetal.2008, Darg2010}, but is comparable to the lower limit of the galaxy pairwise velocity dispersion measured at projected distances of $150 \leq r_p \leq 5000\,h^{-1}_{70}\,\text{kpc}$ \citep{Zehavietal.2002}. 

    We first applied criterion (1) to select 103,216 AGNs with redshifts $z < 0.5$ from the MQC as our initial sample. Based on their redshifts, we calculated the projected distance corresponding to $100\,h_{70}^{-1}$ kpc according to the criterion (2). The initial sample was then self-matched with TOPCAT \citep{topcat2005}, adopting the maximum angular separation as the matching radius. From the self-matched results, we extracted candidate pairs satisfying criterion (2), and subsequently calculated their LOS velocity differences ($|\Delta v|$) following \citet{hogg2000distancecosmos}. Applying criterion (3), we identified 319 dual AGN candidates.
    To exclude previously known systems, we cross-matched these candidates against several comprehensive dual AGN catalogs \citep{Koss2011ApJ, Koss2012ApJL, Liu2011ApJ, Zhang2021AJ, Chen2023ApJ, Pfeifle2025} and also checked the Simbad database\footnote{SIMBAD database: \url{https://simbad.u-strasbg.fr/simbad/}} \citep{simbad2000} for possible duplicates in recent years. This left us with 184 candidate systems. 
    We then visually inspected the optical images of DESI-LS, rejecting six false pairs and identifying five multiplets (four triplets and one quadruplet). Finally, our sample comprises 178 systems: 173 dual AGNs, 4 triplets, and 1 quadruplet. The basic properties of all systems are summarized in Table \ref{catalog_sys}.

\begin{deluxetable*}{lrrrrrrrrrrrllc}
\tablecaption{Catalog of dual AGNs in this work (system-based format).\label{catalog_sys}}
\tablewidth{\textwidth}
\tablehead{
    \colhead{\textbf{Obj. Name}} & 
    \colhead{\textbf{RA(A)}} & 
    \colhead{\textbf{Dec(A)}} & 
    \colhead{\textbf{RA(B)}} & 
    \colhead{\textbf{Dec(B)}} & 
    \colhead{\textbf{$z_A$}} & 
    \colhead{\textbf{$z_B$}} & 
    \colhead{\textbf{$\Delta \theta$}} & 
    \colhead{\textbf{$r_p$}} & 
    \colhead{\textbf{$\Delta v$ }} & 
    \colhead{\textbf{$r_A$}} & 
    \colhead{\textbf{$r_B$}} & 
    \colhead{\textbf{Class(A,B)}} & 
    \colhead{\textbf{Cite}} &
    \colhead{\textbf{$F_{tidal}$}} \\
    \colhead{} &
    \colhead{(J2000)} & 
    \colhead{(J2000)} & 
    \colhead{(J2000)} & 
    \colhead{(J2000)} & 
    \colhead{} & 
    \colhead{} & 
    \colhead{($^{\prime\prime}$)} & 
    \colhead{($h^{-1}_{70}$ kpc)} & 
    \colhead{(km s$^{-1}$)} &
    \colhead{(mag)} &
    \colhead{(mag)}
}
\startdata
J0018$-$0045 & 4.51177 & $-$0.75948 & 4.51300 & $-$0.74598 & 0.0636 & 0.0640 & 48.79 & 59.74 & 98.96 & 17.18 & 17.10 & N, A & DR16, 6dF & N \\
J0103+0029 & 15.81653 & 0.48763 & 15.81621 & 0.48838 & 0.2021 & 0.2025 & 2.94 & 9.77 & 110.79 & 20.19 & 17.89 & N, N & DR16, DR16 & T \\
J0113$-$1450 & 18.45865 & $-$14.84566 & 18.45771 & $-$14.84919 & 0.0520 & 0.0540 & 13.15 & 13.34 & 569.80 & 10.09 & 14.03 & ARX, N & 1237, 6dF & T \\
J0114$-$5523 & 18.59292 & $-$55.39700 & 18.60388 & $-$55.39705 & 0.0120 & 0.0122 & 22.42 & 5.51 & 58.16 & 8.41 & 12.67 & N, NRX & 930, 6dF & T \\
J0120$-$0829 & 20.19541 & $-$8.49063 & 20.20007 & $-$8.48844 & 0.0340 & 0.0344 & 18.36 & 12.45 & 110.23 & 13.15 & 12.53 & N, NRX & PGC, PGC & T \\
J0120$-$4407 & 20.04900 & $-$44.13282 & 20.08202 & $-$44.12858 & 0.0221 & 0.0234 & 86.65 & 38.79 & 372.89 & 10.25 & 6.70 & NX, NRX & PGC, PGC & T \\
J0457$-$7527 & 74.42021 & $-$75.45917 & 74.34222 & $-$75.44512 & 0.0179 & 0.0190 & 86.79 & 30.69 & 473.43 & 6.74 & 4.94 & NX, N & PGC, PGC & N \\
J0852+4121 & 133.24919 & 41.36016 & 133.24631 & 41.36751 & 0.1342 & 0.1348 & 27.68 & 65.66 & 149.67 & 15.63 & 16.81 & AR, AR2 & DR16, DR16 & N \\
J1014+0327 & 153.56228 & 3.46658 & 153.44030 & 3.42467 & 0.0030 & 0.0040 & 463.57 & 28.77 & 298.95 & 5.88 & 5.8 & ARX, NRX & 1592, PGC & N \\
J1612+2934 & 243.04692 & 29.57412 & 243.06965 & 29.57306 & 0.0536 & 0.0542 & 71.26 & 74.41 & 165.10 & 8.83 & 12.81 & ARX, NRX & PGC, PGC & N \\
J2254+0046 & 343.71762 & 0.77538 & 343.71429 & 0.77648 & 0.0908 & 0.0913 & 12.64 & 21.39 & 123.74 & 16.82 & 15.68 & QR, N & DR16, DPeake & T \\
J2353$-$3036 & 358.49433 & $-$30.60231 & 358.49467 & $-$30.60008 & 0.3134 & 0.3144 & 8.06 & 37.02 & 228.33 & 19.28 & 19.52 & QR, N & 2QZ, 2QZ & N \\
\enddata
\tablecomments{Columns: (1) System name; (2–5) Right ascension and declination (degrees) of components A and B; (6–7) Spectroscopic redshifts of A and B; (8–10) Angular separation, projected distance, and the LOS velocity difference; (11–12) $r$-band magnitude of A and B; (13) Classifications given by the MQC, see Section~\ref{subsec21} for details
; (14) Reference to the original discoveries, cite numbers correspond to the Milliquas reference list\footnote{\url{https://www.quasars.org/milliquas-references.txt}}. And most of these sources originate from small-scale follow-up observations with non-survey telescopes; (15) Tidal feature flag ("T" = tidal pair, "N" = non-tidal pair).\\
(This entire table is available in machine-readable form.)}
\end{deluxetable*}

\subsection{Dual AGNs with Tidal Features} \label{subsec23}

\noindent
    We then derived a subset of dual AGN candidates residing in galaxy mergers unambiguously experiencing strong tidal encounters. The selection was made through visual inspection of DESI-LS optical images (DR9 \textit{grz} and DR10 \textit{griz}), focusing on systems exhibiting tidal features such as bridges, tails, shells, or rings. 
    We use visual examination for several reasons:
    (1) Visual inspection is more sensitive to low surface brightness (LSB) features such as bridges, tails, shells, and rings. These features are often critical for identifying tidal interactions but can be challenging to detect with automated techniques. (2) Compared to automated techniques like model fitting, visual identification is less prone to false positives. This means that we are less likely to misidentify misleading or elusive features as real tidal interactions. (3) While quantitative measures of mergers, such as those by \citet{Conselice2003} and \citet{Lotzetal.2004}, are more objective and can produce reproducible results, they may introduce biases towards certain merger stages. For instance, these measures may be susceptible to the first pericenter passage or the final coalescence of merging galaxies, as noted by \citet{Lotzetal.2008}. Visual identification helps mitigate these biases by providing a more comprehensive assessment of tidal features across different stages of the merger process \citep{Liu2011ApJ}.
    To reduce the uncertainty of the visual inspection, two of our members (Z.J.D. and C.X.) independently inspected each pair three times, without knowing the results from previous identifications.  98\% of the trial results were consistent, for objects with different classifications, the category assigned in the majority of the trials (two out of three occurrences) was adopted as the final classification.

 \begin{figure*}
   \centering
   \includegraphics[width=1\linewidth]{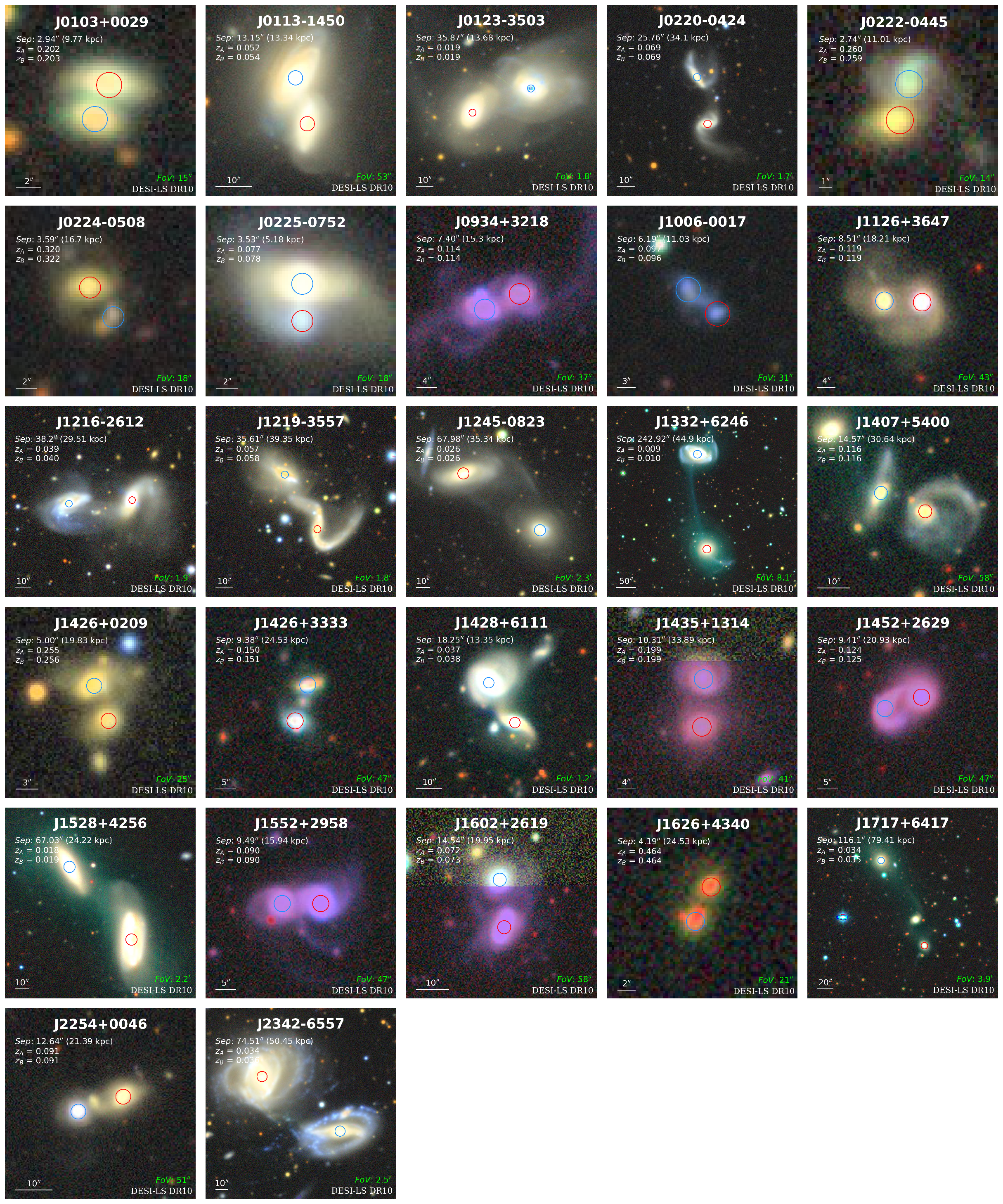}
   \caption{ DESI$-$LS \textit{grz(i)} images of each tidal system, showing clear tidal features associated with dual AGNs. Red and blue circles mark the locations of the AGNs. North is up and east is to the left. The field of view (FOV) of each image is labeled at the bottom right corner. The redshifts, $r_p$, and angular separation of each pair are also presented. The pinkish appearance in some images is likely due to incomplete band coverage.}
   \label{tidal}
 \end{figure*}

    We identified 27 dual AGNs with tidal features in the optical images.
    Figure \ref{tidal} shows representative examples of these systems, where the optical morphologies display prominent tidal tails, bridges, or distorted isophotes suggestive of ongoing galaxy interactions. In some systems (e.g., J0103+0029, J1426+0209, J1626+4340), only marginally extended or asymmetric structures are visible, making it difficult to identify more prominent bridge- or ring-like features. We also look forward to obtaining more images at declinations below $-30^\circ$ in the future to identify additional tidal systems.

\section{Sample Statistics and Results} \label{sec3}
    Our systematic search and analysis in the MQC has led to the identification of 178 new multi-AGN systems at low redshifts ($z < 0.5$), which include 173 pairs, 4 triplets, and 1 quadruplet, all with $r_p < 100h_{70}^{-1}$ kpc and $|\Delta v| < 600$ km s$^{-1}$. Here we present some results and statistics on these AGN systems. The optical images of the dual AGNs are presented in the Appendix~\ref{appendix}. For each system, three panels are shown from left to right: the DESI–LS DR10 image, the DESI–LS DR10 model image, and the DESI–LS DR10 residual image. These images provide a clear visual representation of the morphological characteristics of the sources in our sample and the results of the model fitting.

\subsection{Statistics and Analysis of Dual AGNs} \label{subsec31}
    MQC is an inhomogeneous catalog that compiles AGNs from multiple surveys with varying target selection criteria, spectral quality, and data completeness. Such differences may influence our statistical analyses, similar to the systematic band- and technique-dependent biases emphasized in the Big MAC DR1 study (\citet{Pfeifle2025}, Section 4). 

\begin{figure}
    \centering
    \includegraphics[width=1\linewidth]{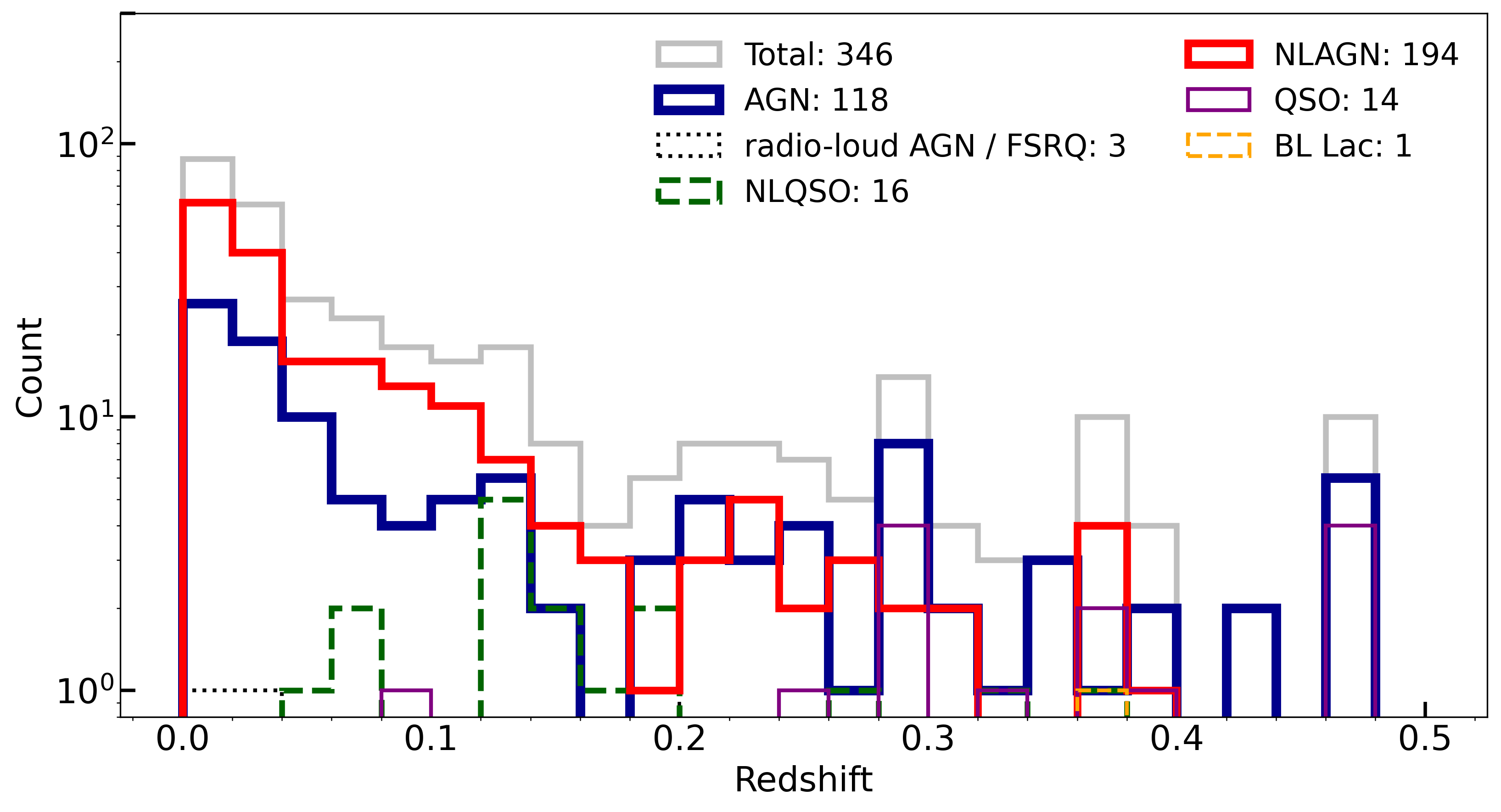}
    \caption{Redshift distribution of various types of sources in dual AGN sample.}  \label{z_dis}
   \end{figure}

    Our data consist of 346 individual AGNs within 173 dual systems. Figure \ref{z_dis} shows the redshift distribution of various source types in our dual AGN sample. This distribution suggests that, among low-redshift dual AGN systems in the MQC, the majority are low-luminosity galaxies, predominantly narrow-line AGNs (56.1\%) and Seyfert-type broad-line AGNs (34.1\%). In contrast, QSOs, BL Lacs, and radio-loud AGNs/FSRQs are relatively rare, together comprising only about 5\%. Table~\ref{tab:count} provides detailed statistics on the classification combinations in our dual AGN samples.
    BL Lac objects and broad-line radio-loud AGNs/FSRQs are combined into a single 'blazar' category for statistical purposes, following the conventional classification of blazars \citep{Urry1995}. The results show that AGN-NLAGN (35.3\%) and NLAGN-NLAGN (35.8\%) pairs dominate our sample, together accounting for more than 70\% of all dual AGNs. This indicates that narrow-line AGNs make up the majority of low-redshift dual AGN systems. Such a trend is broadly consistent with the results of \citet{Liu2011ApJ} on dual AGNs, who reported that narrow-line AGNs constitute up to 96\% of their sample. In addition, the AGN-AGN combination of broad-line Seyfert AGNs also represents a significant fraction (14.5\%). In contrast, quasar combinations (e.g., QSO-QSO at 1.7\%, AGN-QSO at 2.9\%, and NLAGN-QSO at 1.7\%) are relatively rare, likely reflecting the scarcity of dual quasars at low redshifts. In addition, combinations with blazars or blazar-NLQSO are extremely rare, together accounting for only $\sim$1–2\% of the sample.
            
   \begin{deluxetable*}{lcr lcr lcr}
    \tablecaption{Statistical analysis of classification combinations in the dual AGN sample\label{tab:count}}
    \tablewidth{\textwidth}
    \tablecolumns{8}
    \tablehead{
    \colhead{combination} & \colhead{Count} & \colhead{\%} &
    \colhead{combination} & \colhead{Count} & \colhead{\%} &
    \colhead{combination} & \colhead{Count} & \colhead{\%}
    }
    \startdata
    AGN-NLAGN  & 61 & 35.3 &     NLAGN-NLAGN & 62 & 35.8 &     NLQSO-AGN  & 2 & 1.2 \\
    AGN-AGN  & 25 & 14.5 &     NLQSO-NLQSO  & 5 & 2.9  &    Blazar-NLAGN  & 2 & 1.2 \\
    AGN-QSO  & 5 & 2.9 &    NLQSO-NLAGN &  4 & 2.3 &     Blazar-Blazar & 1 & 0.6  \\
    QSO-QSO  & 3 & 1.7 &    NLAGN-QSO   &  3 & 1.7 &      &  \\
    \enddata
    \tablecomments{Percentages are computed with respect to the full dual-AGN sample ($N=173$).}
   \end{deluxetable*} 
   
    Moreover, we identified 27 pairs with tidal features in optical images, corresponding to approximately 16\% of our AGN sample. A subset of these systems is marked with a “T” flag in Table \ref{catalog_sys} to explicitly indicate their tidal morphology. 
    As a comparison, \cite{Liu2011ApJ} visually inspected SDSS images and found that $\sim$30\% of their 1286 pairs showed tidal features, markedly higher than our result. It is a bit surprising that we found a much lower proportion of systems with tidal features despite having much deeper images.
    The tidal sample may lack mergers in their earliest stages prior to the first pericenter passage, as such events lack observable tidal features \citep{Toomre&Toomre1972}. Furthermore, the strength of tidal signatures is influenced by the properties of the host galaxies and orbital parameters of the progenitor systems.
    While gas-rich major mergers on prograde orbits may readily generate strong tidal features \citep{Barnes1996, Cox2008}, other merger populations lacking discernible optical tidal features may consist of minor mergers with mass ratios exceeding 30 \citep{Lotzetal.2010}, spheroidal mergers characterized by high bulge-to-disk ratios \citep{Mihos&Hernquist1996}, and mergers occurring on highly retrograde orbits \citep{Toomre&Toomre1972}.
   
\begin{figure}
     \centering
     \includegraphics[width=1\linewidth]{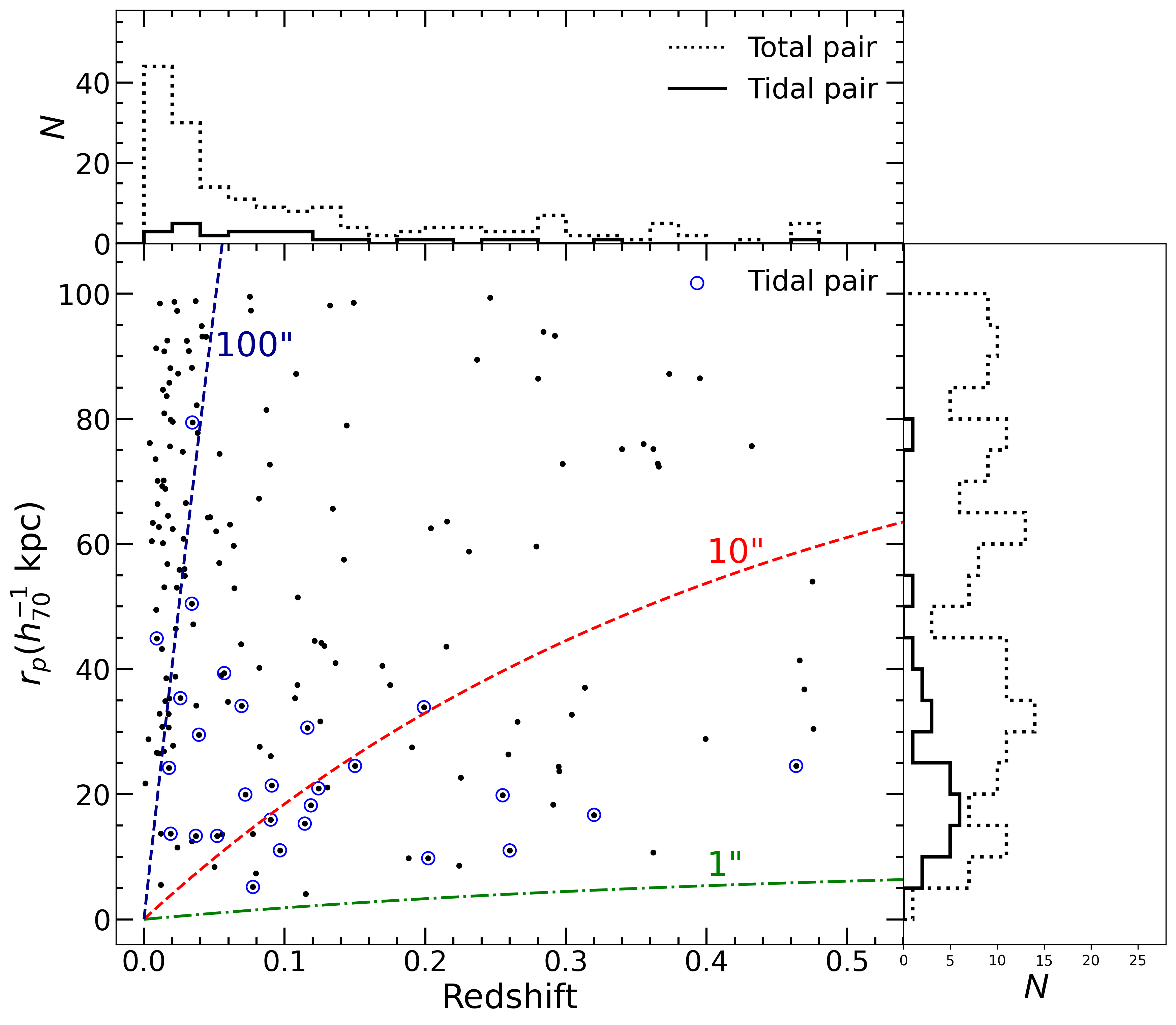}
     \caption{Projected distance ($r_p$) vs. redshift for our dual AGNs. Black dots represent all dual AGNs, while blue circles represent dual AGNs with identified morphological tidal features identified in the optical images. The dotted histograms represent the distributions of the full dual AGN sample, while the solid histograms show the distributions of the tidal sample. The green, red, and blue dashed lines correspond to angular separations of 1$^{\prime\prime}$, 10$^{\prime\prime}$, and 100$^{\prime\prime}$, respectively.}  \label{z_rp}
   \end{figure}

    Figure \ref{z_rp} shows the distribution of projected distance versus redshift for our sample. The number of dual AGNs increases toward lower redshifts. This redshift distribution is very likely a systematic effect driven by the selection methodology for MQC samples. For projected distance, the number is more evenly distributed. But in some bins, such as 45 $\leq r_p \leq 50 \, h^{-1}_{70} \, \text{kpc}$, contain relatively few systems, which may be due to the inhomogeneity of the parent sample. Similarly, the number of systems within $0 \leq r_p\leq 5 \, h^{-1}_{70} \, \text{kpc}$ is relatively lower, likely reflecting the angular resolution limits of the surveys. The tidal subsample is mainly concentrated at $z < 0.14$, and have separations in the range of $0-55\, h^{-1}_{70}$kpc. As $r_p$ decreases, the number of tidal systems apparently increases, peaking at $10-20\,h^{-1}_{70}$ kpc.
 

\begin{figure}
        \centering
        \includegraphics[width=1\linewidth]{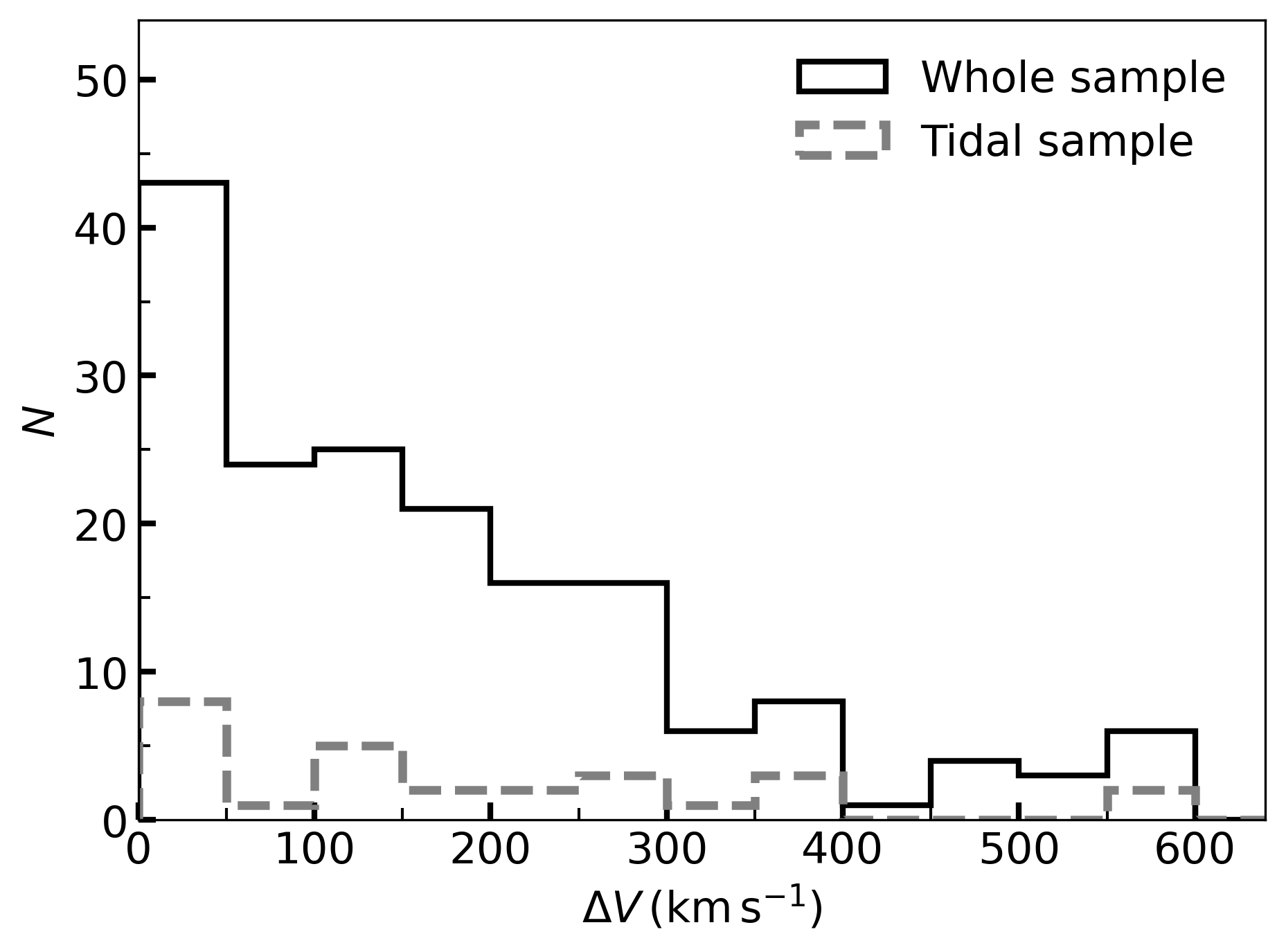}
        \caption{Distribution of LOS velocity difference $|\Delta v|$ for our dual AGNs. The black histogram represents the distribution of all dual AGNs, while the dark blue dashed line shows the distribution of the tidal sample.}
        \label{deltV_dis}
    \end{figure}

    Figure~\ref{deltV_dis} shows the distribution of the LOS velocity differences $|\Delta v|$ for our dual AGNs. The distribution shows a decreasing trend with increasing $|\Delta v|$. When $|\Delta v| > 400\,\text{km s}^{-1}$, the full and tidal samples contain only a small number of systems (15 dual AGNs in total, of which 3 are tidal systems). 
    Notably, there is a drop-off in the number of pairs with $|\Delta v| > 300\,\text{km s}^{-1}$. The selection effect of the MQC might play some sort of role in the behavior of the distribution, this drop is consistent with that seen by \citep{Liu2011ApJ}.
    For the tidal subsample, no clear trend is observed due to the limited number of systems.

\subsection{Close- and Wide-Separation Systems}
\begin{figure}
       \centering
       \includegraphics[width=1\linewidth]{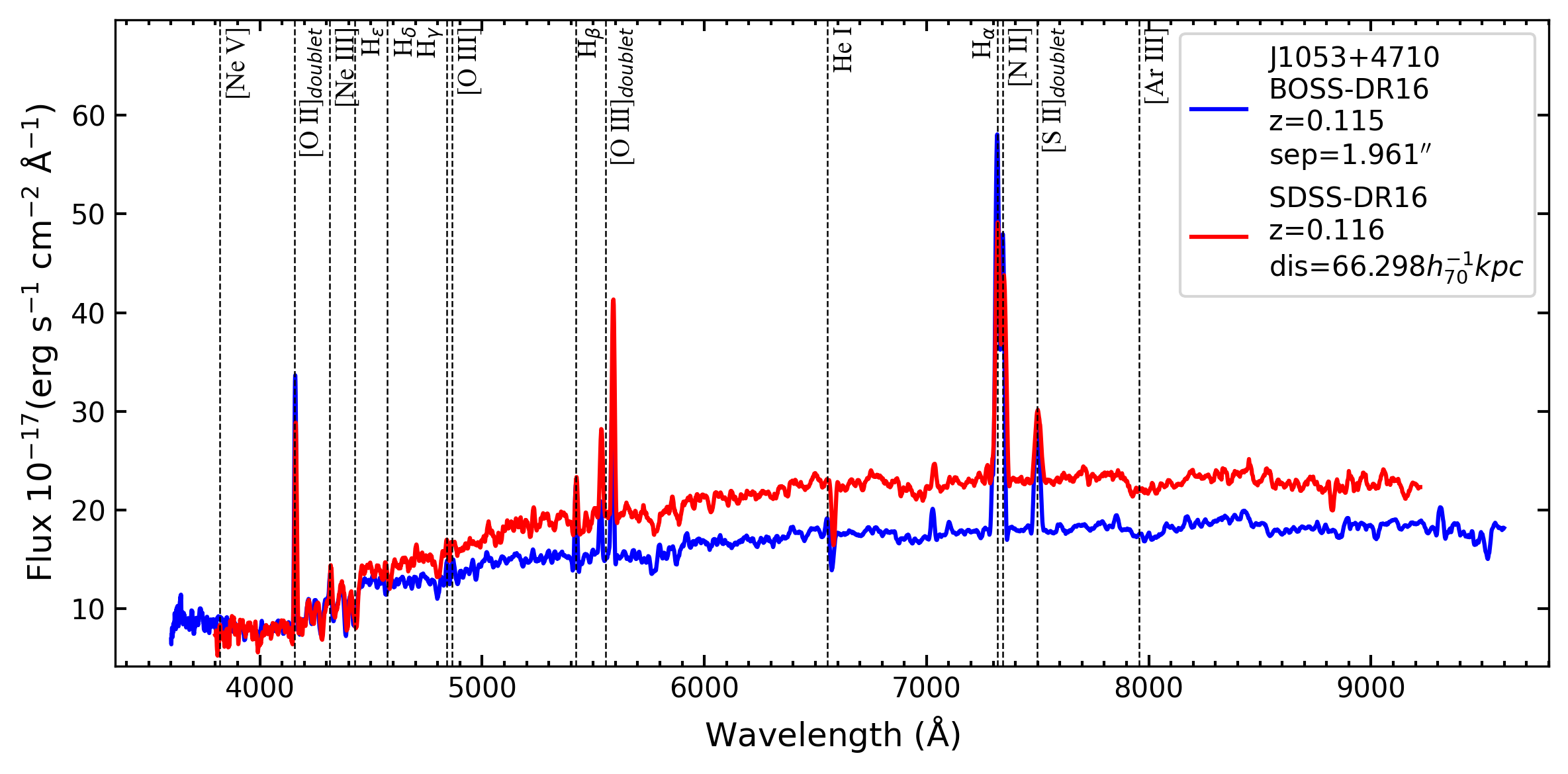}
       \caption{Comparison of SDSS and BOSS spectra for the dual AGN system J1053+4710. Major emission lines are marked with vertical dashed lines, and the basic properties of the two components are shown in the upper-right corner.}
       \label{small}
    \end{figure}

    At small angular separations, we identified a dual AGN candidate, J1053+4710, with a projected distance of $r_p < 5\, h^{-1}_{70}$ kpc and an angular separation of 1.961$^{\prime\prime}$. The residual image shown in Appendix \ref{appendix} further reveals two prominent nuclei. However, the emission line strengths and continuum levels of the two spectra appear similar in Figure~\ref{small}. A more critical issue for such close pairs drawn from fiber surveys is fiber spillover. 
    For our sample, close angular pairs with separations $< 10^{\prime\prime}$ may suffer from such spillover effects, meaning some emission from both nuclei could be contained in both fibers. This complicates the definitive spectroscopic confirmation based on the survey data alone. Therefore, these close pairs, including J1053+4710, should be considered as dual AGN candidates and require spatially resolved follow-up spectroscopy for verification.
    
\begin{figure}
        \centering
        \includegraphics[width=1\linewidth]{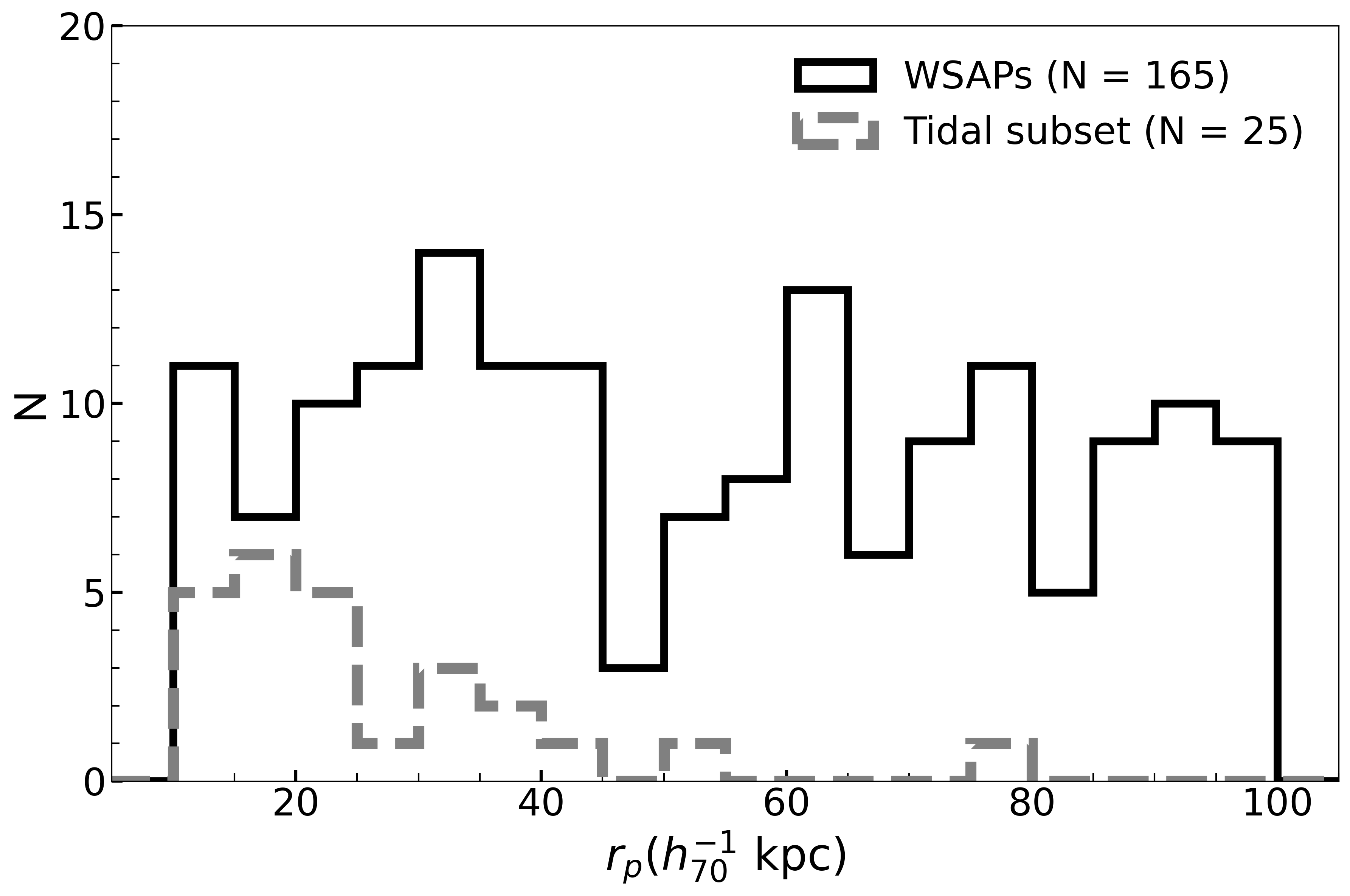}
        \caption{Distribution of projected distance ($r_p$) for our dual AGNs. The black solid step histogram represents the full WSAP sample (N=165), while the gray dashed step histogram corresponds to the tidal subset (N=25).}
        \label{rp_dis}
    \end{figure}

    Furthermore, we define those pairs with $r_p >10\, h^{-1}_{70}$ kpc as wide separation dual AGNs (WSAPs). Our sample includes 165 WSAPs, with 25 displaying clear tidal features, which are shown in Figure \ref{tidal}. Figure \ref{rp_dis} illustrates the number of $r_p$ distributions of these WSAPs. 
    The full WSAP sample shows a relatively flat distribution as a function of $r_p$, but the tidal sample drops off relatively quickly, with few tidal pairs out past 40 $h^{-1}_{70}$kpc.


    The WSAPs allow us to observe more detailed interaction features in these systems. 
    They are particularly valuable for tracing the early and intermediate stages of galaxy interactions, when gravitational torques and tidal perturbations begin to drive gas inflows toward the nuclei \citep{Barnes&Hernquist1992, Hopkins2010}.
    The efficiency of such fueling depends on the orbital geometry, mass ratio, and gas content of the progenitors, producing diverse kinematic and morphological signatures \citep{Capelo2017MNRAS, Cox2008}.
    Observational studies have shown that nuclear activity can arise in early-stage mergers and persist out to large separations \citep{Liu2011ApJ, Koss2012ApJL, DeRosa2023MNRAS}.
    Dual AGNs within $<$60 kpc often show gas inflow toward the nuclei, while systems beyond $\sim$100 kpc may evolve without strong gas disturbances \citep{DeRosa2023MNRAS}.
    Together, these results highlight the importance of investigating large-scale (10–100 kpc) dual AGNs, and our WSAP subsample offers an opportunity to probe the early co-evolution of SMBHs and their hosts, thereby enabling extrapolation to higher redshifts and enhancing our understanding of the fine details of merger-induced mass assembly.

\subsection{Multiples, Extra Galaxies and/or AGN Candidates in the Pair Regions}
\begin{figure*}
       \centering
       \includegraphics[width=1\linewidth]{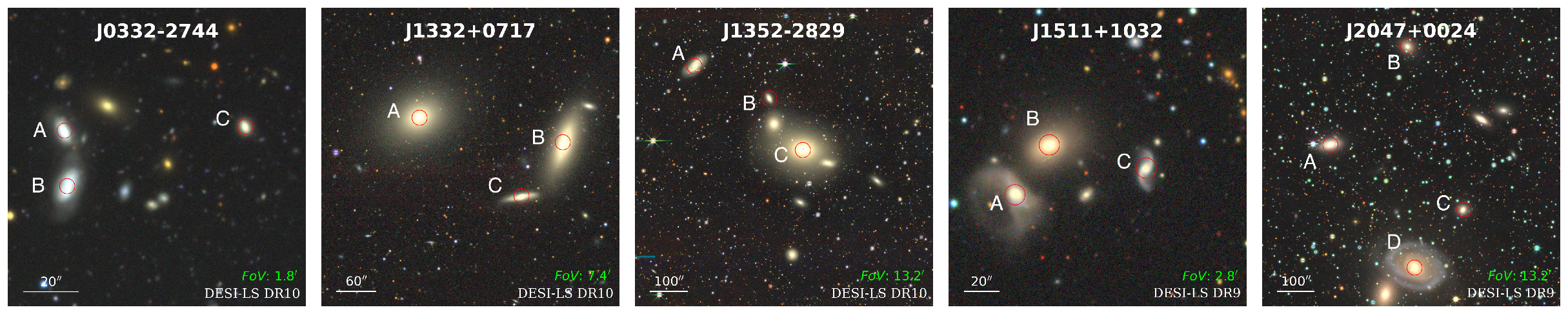}
       \caption{DESI–LS grz(i)-bands cutouts of five multi–AGN systems in our sample: four triple AGNs and one quadruple AGN. The components are labeled as A–D in each field.}
       \label{multi_AGN}
    \end{figure*}

    Our sample includes five multiple AGNs (four triplets and one quadruplet), as presented in Figure \ref{multi_AGN} and listed in Table \ref{multi_para}, where a system is included if at least one pair of objects satisfies the projected distance/ LOS velocity difference criterion outlined in Section \ref{subsec22}.
    Some are optically selected (e.g., from SDSS DR16 or GSAC), while others are drawn from galaxy catalogs (PGC) or literature-based AGN compilations. 
    
    The projected distances among the nuclei in these systems range from 40 to 120$\, h^{-1}_{70}$ kpc. For instance, the closest members in J1332+0717 and J1511+1032 are separated by $\sim40–50\, h^{-1}_{70}$ kpc, while the most wide separation pairs exceed 90–100$\, h^{-1}_{70}$ kpc. In the quadruplet J2047+0024, the distances span from $\sim49 \, h^{-1}_{70}$ kpc to more than 110$\,h^{-1}_{70}$ kpc. Notably, no sub-10 kpc multiple systems are present in our sample, underscoring the extreme rarity of compact triple or quadruple AGNs at low redshifts.
    DESI–LS images reveal disturbed host morphologies and possible tidal-like distortions in J1511+1032, consistent with interaction signatures. However, other systems lack obvious tidal features.
    
    \begin{deluxetable*}{lcrrlrrl}
        \tablecaption{Basic parameters of the multi-AGN systems.\label{multi_para}}
        \tablecolumns{8}
        \tablehead{
            \colhead{\textbf{System}} & 
            \colhead{\textbf{No.}} & 
            \colhead{\textbf{RA}} & 
            \colhead{\textbf{DEC}} & 
            \colhead{\textbf{Class}} & 
            \colhead{\textbf{Redshift}} &  
            \colhead{\textbf{$r_p$}} & 
            \colhead{\textbf{Cite}} \\
            \colhead{} &
            \colhead{} & 
            \colhead{(J2000)} & 
            \colhead{(J2000)} & 
            \colhead{} & 
            \colhead{} &
            \colhead{($h^{-1}_{70}$ kpc)} & 
            \colhead{}
    }
    \startdata
    J0332$-$2744 & A: & 53.124506 & $-$27.740284 & \;AX & 0.076 & 28.36916 & GSAC \\
    & B: & 53.104588 & $-$27.734287 & \;A & 0.076 & 96.60849 & GSAC \\
    & C: & 53.124941 & $-$27.734830 & \;AX & 0.076 & 93.50298 & GSAC \\
    \hline 
    J1332+0717 & A: & 203.134409 & 7.294156 & \;A & 0.023 & 48.04549 & PGC \\
    & B: & 203.116300 & 7.316590 & \;NR & 0.023 & 102.55542 & PGC \\
    & C: & 203.177200 & 7.327290 & \;AR & 0.023 & 90.01932 & PGC \\
    \hline 
    J1352$-$2829 & A: & 208.312600 & $-$28.427610 & \;NX & 0.015 & 65.55664 & PGC \\
    & B: & 208.250722 & $-$28.451806 & \;NX & 0.017 & 52.92251 & PGC \\
    & C: & 208.222200 & $-$28.489320 & \;AX & 0.016 & 110.77492 & PGC \\ 
    \hline 
    J1511+1032 & A: & 227.785414 & 10.537003 & \;NR & 0.07 & 43.55189 & PGC \\
    & B: & 227.780110 & 10.544547 & \;A & 0.069 & 71.37602 & PGC \\
    & C: & 227.764846 & 10.540903 & \;A & 0.067 & 95.18574 & DR16\\
    \hline
    J2047+0024 & A: & 311.892100 & 0.411622 & \;NX & 0.012 & 81.53752 & 0478\\
    & B: & 311.834900 & 0.483990 & \;N & 0.014 & 112.69869 & PGC \\
    & C: & 311.793786 & 0.363305 & \;ARX & 0.012 & 49.02998 & PGC \\
    & D: & 311.829437 & 0.320811 & \;A & 0.014 & 97.52666 & PGC \\
    \enddata
        \tablecomments{Basic parameters of the five multiple AGN systems. The “Type” column gives the classifications provided in the MQC. The $r_p$ column lists the projected distance between the paired components (e.g., A–B, B–C, etc.). The entry "0478" in the cite column is a reference to Milliquas, see its link provided in Table\ref{catalog_sys}.}
    \end{deluxetable*}

    During the visual inspection of our sample, we found several additional galaxies and/or AGN candidates located in the same regions as the pairs or multiplets. Based on the criteria laid out in Section \ref{subsec22}, we identified 14 three-object, 8 four-object, 2 five-object, one six-object, one eight-object, and one nine-object systems. In this process, we also identified one additional LINER not included within the MQC. The basic data for a subset of these additional galaxy multiplets are listed in Table \ref{catalog_2}, along with their classifications obtained from SIMBAD (provided in the “Class” column).  These AGN candidates are worthy of further observational confirmation, and the entire system merits in-depth study at multiple wavelengths.

    \begin{figure*}
       \centering
       \includegraphics[width=1\linewidth]{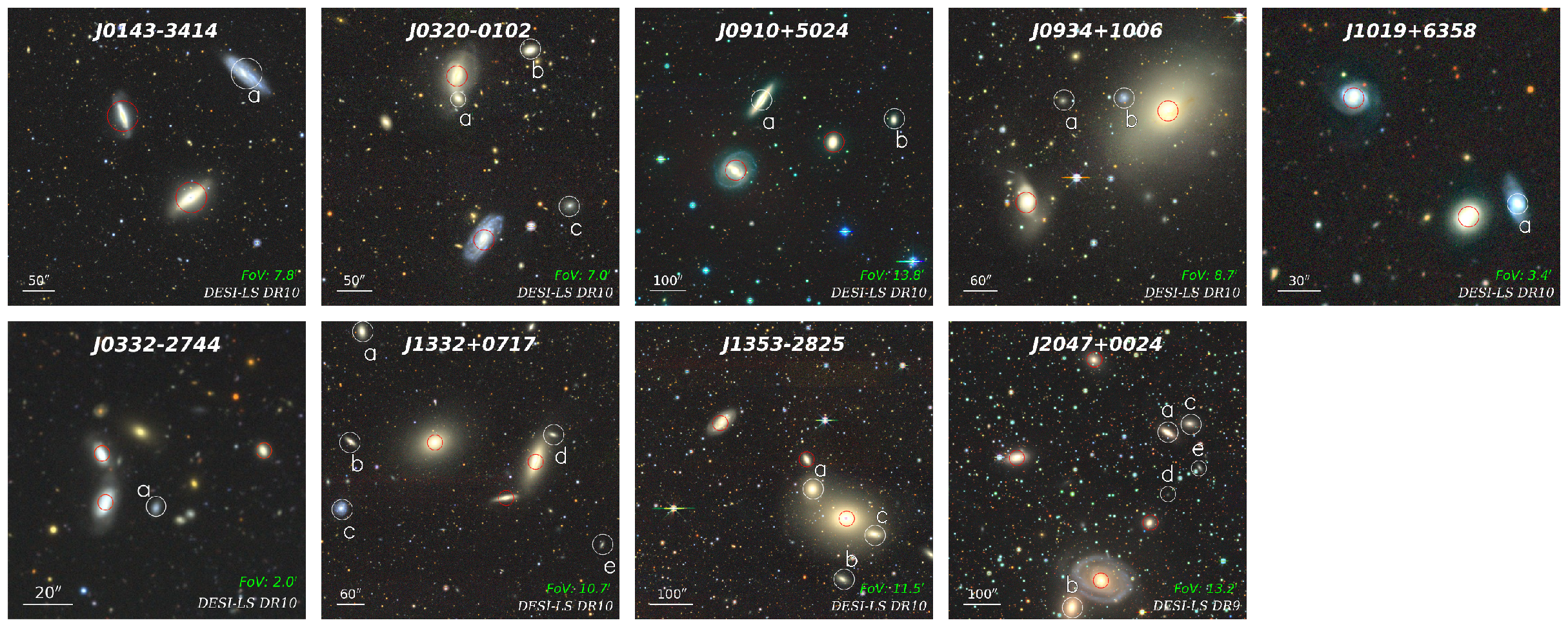}
       \caption{DESI-LS grz(i)-bands images of five pairs, three triplets, and a quadruplet with extra galaxies and/or AGN candidates. Red circles indicate the locations of AGNs. The numbered extra galaxies and/or AGN candidates are in white circles. North is up and east is to the left. The FOV of each image is labeled at the bottom right corner of the image. }
       \label{extra}
    \end{figure*}
    
\begin{deluxetable*}{clrrrrrrl}
\tablecaption{Catalog of Extra galaxies or AGNs (or Candidates) in the pair regions. \label{catalog_2}}
\tablewidth{\textwidth}
\tablecolumns{9}
\tiny
\tablehead{
    \colhead{\textbf{System}} & 
    \colhead{\textbf{Obj.name}} &
    \colhead{\textbf{RA}} & 
    \colhead{\textbf{Dec}} & 
    \colhead{\textbf{\textit{z}}} & 
    \colhead{\textbf{$\Delta \theta$}} & 
    \colhead{\textbf{$r_p$}} & 
    \colhead{\textbf{$|\Delta v|$}} & 
    \colhead{\textbf{Class}} \\
    \multicolumn{1}{c}{} & \multicolumn{1}{c}{} & \multicolumn{1}{c}{(J2000)} & \multicolumn{1}{c}{(J2000)} & 
    \multicolumn{1}{c}{} & \multicolumn{1}{c}{(\arcsec)} & \multicolumn{1}{c}{($h^{-1}_{70}$ kpc)} & 
    \multicolumn{1}{c}{(km s$^{-1}$)} & \multicolumn{1}{c}{} \\
   \multicolumn{1}{c}{(1)} & \multicolumn{1}{c}{(2)} & \multicolumn{1}{c}{(3)} & 
    \multicolumn{1}{c}{(4)} & \multicolumn{1}{c}{(5)} & \multicolumn{1}{c}{(6)} & 
    \multicolumn{1}{c}{(7)} & \multicolumn{1}{c}{(8)} & \multicolumn{1}{c}{(9)} 
}
\startdata
\scriptsize J0143-3414 & \scriptsize\textcircled{a}IC 1722 & \scriptsize25.76242 & \scriptsize-34.18733 & \scriptsize0.01384 & \scriptsize202.89 & \scriptsize52.184 & \scriptsize371.89 & \scriptsize \;LSBG \\
\hline
\scriptsize \multirow{3}{*}{J0320-0102} & \scriptsize\textcircled{a}LEDA 12538
& \scriptsize50.18886 & \scriptsize-1.05395 & \scriptsize0.02079 & \scriptsize37.51 & \scriptsize15.783 & \scriptsize179.22 & \scriptsize \;Galaxy \\
& \scriptsize\textcircled{b}Z 390-67 & \scriptsize 50.16065 & \scriptsize-1.03503 & \scriptsize0.02101 & \scriptsize112.52 & \scriptsize47.839 & \scriptsize114.57 & \scriptsize \;EmG \\
& \scriptsize\textcircled{c}LEDA 1126869 & \scriptsize50.14511 & \scriptsize-1.09587 & \scriptsize0.02139 & \scriptsize133.33 & \scriptsize56.433 & \scriptsize141.02 & \scriptsize \;EmG \\
\hline
\scriptsize\multirow{2}{*}{J0910+5024} & \scriptsize\textcircled{a}NGC 2769
& \scriptsize137.63393 & \scriptsize50.43325 & \scriptsize0.01607 & \scriptsize193.81 & \scriptsize63.411 & \scriptsize274.46 & \scriptsize \;LINER \\
& \scriptsize\textcircled{b}NGC 2762 & \scriptsize137.47723 & \scriptsize50.41826 & \scriptsize0.01564 & \scriptsize179.25 & \scriptsize57.111 & \scriptsize233.26 & \scriptsize \;AGN Candidate \\
\hline
\scriptsize\multirow{2}{*}{J0934+1006} & \scriptsize\textcircled{a}SDSS J093358.56+100928.3 & \scriptsize143.49403 & \scriptsize10.15787 & \scriptsize0.01195 & \scriptsize186.36 & \scriptsize41.168 & \scriptsize347.65 & \scriptsize \;AGN Candidate \\
& \scriptsize\textcircled{b}NGC 2912 & \scriptsize143.46364 & \scriptsize10.15937 & \scriptsize0.01106 & \scriptsize81.82 & \scriptsize18.074 & \scriptsize81.61 & \scriptsize \;AGN Candidate \\
\hline
\scriptsize J1019+6358 & \scriptsize\textcircled{a}MCG+11-13-017 & \scriptsize154.78098 & \scriptsize63.96993 & \scriptsize0.04152 & \scriptsize35.27 & \scriptsize28.919 & \scriptsize0.58 & \scriptsize \;Galaxy \\
\hline
\scriptsize J0332-2744 & \scriptsize\textcircled{a}EIS-DEEP CDFS-1 J 1 247 & \scriptsize53.11808 & \scriptsize-27.74062 & \scriptsize0.0726 & \scriptsize21.05 & \scriptsize29.093 & \scriptsize949.46 & \scriptsize \;EmG \\
\hline
\scriptsize\multirow{5}{*}{J1332+0717} & \scriptsize\textcircled{a}Z 45-12 & \scriptsize203.22101 & \scriptsize7.39347 & \scriptsize0.02251 & \scriptsize285.00 & \scriptsize129.611 & \scriptsize255.15 & \scriptsize \;GinGroup \\
& \scriptsize\textcircled{b}Z 45-11 & \scriptsize203.22781 & \scriptsize7.32769 & \scriptsize0.02375 & \scriptsize180.71 & \scriptsize85.269 & \scriptsize108.44 & \scriptsize \;Galaxy \\
& \scriptsize\textcircled{c}NGC 5212 & \scriptsize203.23345 & \scriptsize7.28786 & \scriptsize0.02375 & \scriptsize245.94 & \scriptsize116.048 & \scriptsize553.54 & \scriptsize \;GinGroup \\
& \scriptsize\textcircled{d}SDSS J133225.48+071954.6 & \scriptsize203.10618 & \scriptsize7.33184 & \scriptsize0.02335 & \scriptsize65.72 & \scriptsize30.971 & \scriptsize8.79 & \scriptsize \;Galaxy \\
& \scriptsize\textcircled{e}SDSS J133218.37+071557.9 & \scriptsize203.07646 & \scriptsize7.26614 & \scriptsize0.02161 & \scriptsize230.70 & \scriptsize100.853 & \scriptsize517.94 & \scriptsize \;AGN Candidate \\
\hline
\scriptsize\multirow{3}{*}{J1353+2825} & \scriptsize\textcircled{a}NGC 5330 & \scriptsize208.24666 & \scriptsize-28.47072 & \scriptsize0.01630 & \scriptsize69.32 & \scriptsize22.997 & \scriptsize101.89 & \scriptsize \;GinGroup \\
& \scriptsize\textcircled{b}2MASX J13525393-2831421 & \scriptsize208.22473 & \scriptsize-28.52837 & \scriptsize0.01708 & \scriptsize140.98 & \scriptsize45.351 & \scriptsize380.83 & \scriptsize \;GinCl \\
& \scriptsize\textcircled{c}LEDA 3094716 & \scriptsize208.20170 & \scriptsize-28.49947 & \scriptsize0.01528 & \scriptsize74.35 & \scriptsize23.158 & \scriptsize150.68 & \scriptsize \;EmG \\
\hline
\scriptsize\multirow{5}{*}{J2047+0024} & \scriptsize\textcircled{a}NGC 6959 & \scriptsize311.78017 & \scriptsize0.43017 & \scriptsize0.01222 & \scriptsize245.67 & \scriptsize61.412 & \scriptsize53.34 & \scriptsize \;Galaxy \\
& \scriptsize\textcircled{b}NGC 6964 & \scriptsize311.85126 & \scriptsize0.30083 & \scriptsize0.01265 & \scriptsize106.52 & \scriptsize27.550 & \scriptsize402.63 & \scriptsize \;AGN Candidate \\
& \scriptsize\textcircled{c}LEDA 162626 & \scriptsize311.76379 & \scriptsize0.43679 & \scriptsize0.01257 & \scriptsize285.74 & \scriptsize72.464 & \scriptsize50.37 & \scriptsize \;AGN Candidate  \\
& \scriptsize\textcircled{d}SDSS J204707.21+002302.9 & \scriptsize311.78001 & \scriptsize0.38418 & \scriptsize0.01284 & \scriptsize90.03 & \scriptsize22.832 & \scriptsize130.35 & \scriptsize \;AGN Candidate \\
& \scriptsize\textcircled{e}LEDA 1165076 & \scriptsize311.75643 & \scriptsize0.40443 & \scriptsize0.01400 & \scriptsize199.99 & \scriptsize50.718 & \scriptsize460.55 & \scriptsize \;AGN Candidate \\
\hline
\enddata
\tablecomments{Columns: (1)-(5) System Name, Name of extra galaxies and/or AGN Candidates, Coordinates (deg), Redshift of extra galaxies or AGNs (or Candidates); (6)-(8) Angular separation, projected distance, and LOS velocity difference from the nearest AGN of the system; (9) Source classifications as retrieved from the SIMBAD database\footnote{\url{https://simbad.cds.unistra.fr/simbad/}}. The abbreviations used are: LSBG for Low Surface Brightness Galaxy, EmG for Emission-line Galaxy, GinGroup for Galaxy towards a Group of Galaxies, and GinCl for Galaxy towards a Cluster of Galaxies.\\ 
(This entire table is available in machine-readable form.)
}
\end{deluxetable*}

\section{Discussion} \label{sec4}
    We performed a systematic search of 103,216 AGNs at low redshifts ($z < 0.5$) in the MQC and identified 178 new multiple AGN systems, comprising 173 dual AGNs, 4 triple AGNs, and 1 quadruple AGN. In the MQC, dual AGNs constitute $\sim$0.34\% of the sample, while triple and quadruple AGNs account for $\sim$0.016\%. For comparison, \citet{Liu2011ApJ} analyzed 138,070 AGNs in the redshift range $0.02 < z < 0.33$ and reported 1,286 multiple AGN systems, including 1,244 dual AGNs, 39 triple AGNs, 2 quadruple AGNs, and 1 quintuple AGN, corresponding to fractions of $\sim$1.8\% and $\sim$0.09\% for dual and triple nucleus systems, respectively. 
    The fraction of multiple AGNs reported by \citet{Liu2011ApJ} is significantly higher than that found in the MQC sample.
    This discrepancy likely reflects the inhomogeneous nature of the MQC and differences in sample composition, most notably in the fraction of narrow-line AGNs (NLAGNs-NLQSOs), which constitute $\sim$61\% in the MQC compared to 94\% in \cite{Liu2011ApJ} sample. In addition, there is a marked concentration of systems at $z < 0.25$. Cosmological simulations predict that the fraction of dual AGNs should increase with redshift and peak around cosmic noon \citep{Chen2020, Foord2024}, suggesting that the observed low-redshift concentration is primarily driven by selection effects inherent to the MQC.
    However, our sample has enhanced coverage at higher redshifts with 25.4\% of pairs at $z>0.16$ compared to only 1.9\% in \citet{Liu2011ApJ}. While the MQC is an inhomogeneous parent sample, we do recover a significant number of higher redshift systems relative to prior samples.

\begin{figure}
      \centering
      \includegraphics[width=1\linewidth]{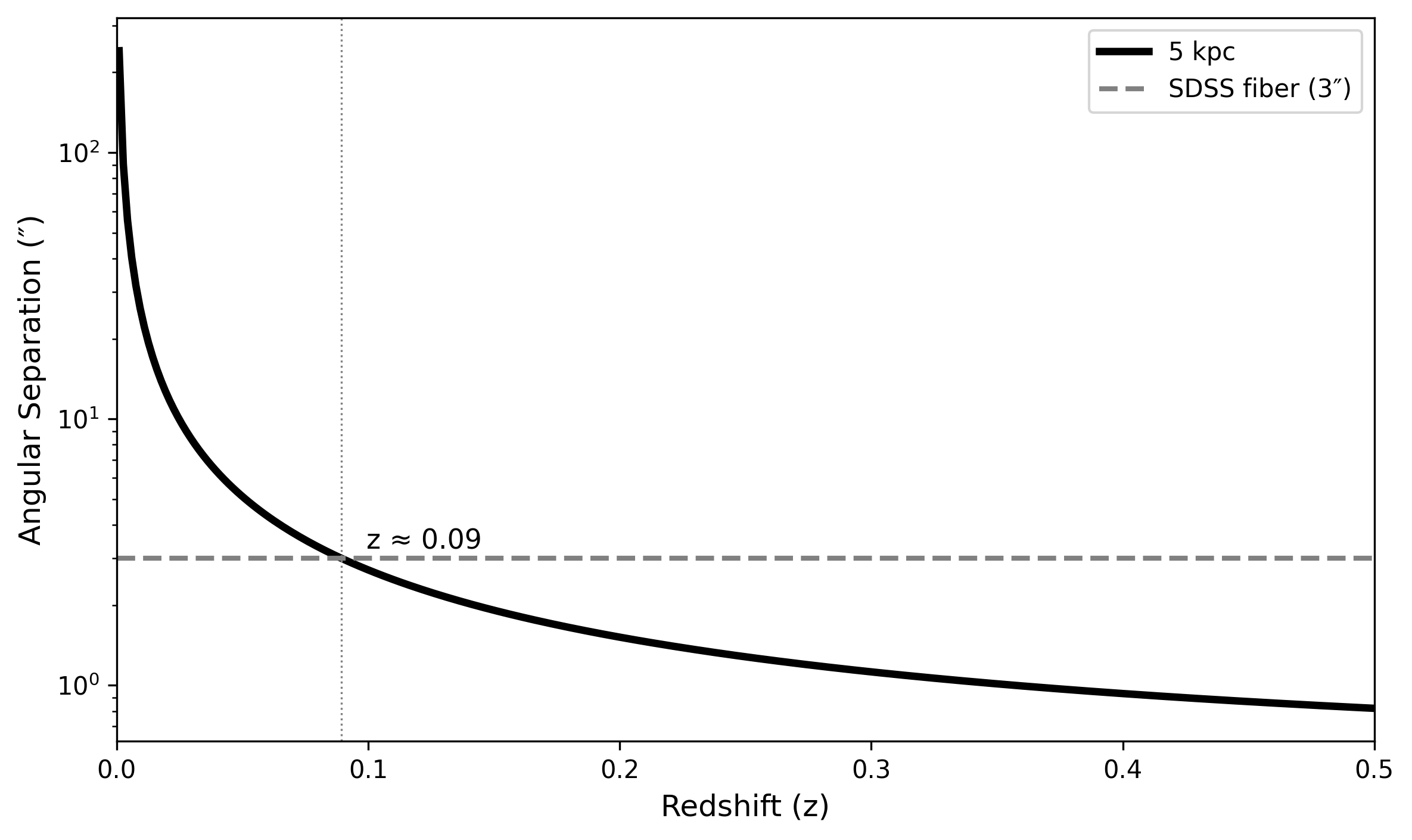}
      \caption{Angular separation as a function of redshift corresponding to a projected distance of 5 h$^{-1}_{70}$ kpc. The blue dashed line indicates the SDSS fiber diameter of 3$^{\prime\prime}$. }
      \label{sep_5kpc}
    \end{figure}

    Second, the dual AGNs are broadly uniformly distributed within the projected distance range of $0–100\,h^{-1}_{70}$kpc, and the tidal subsample is predominantly concentrated at $5–20 \,h^{-1}_{70} $kpc. 
    This is consistent with the statistics of \citet{Koss2012ApJL}, who found that among 167 dual AGNs at $z<0.05$ the majority lie within $<30$ kpc with 50\% (8/16) located at $<15$ kpc. 
    In addition, certain bins of the projected distance distribution (e.g., $45 \leq r_p \leq 50 \, h^{-1}_{70}$ kpc) contain relatively few systems.
    The apparent deficit of systems at small separations ($r_p \leq 5\,h^{-1}_{70}$ kpc) is likely due to the angular resolution limits of major spectroscopic surveys.
    As shown in Figure \ref{sep_5kpc}, the angular separation for $r_p < 5,h^{-1}_{70}$ kpc decreases with increasing redshift within $0 < z < 0.5$, reaching about $3^{\prime\prime}$ at $z \approx 0.09$, comparable to the SDSS fiber size (DESI$\sim 1.5^{\prime\prime}$).
    Therefore, close dual nuclei are likely to be missed \citep{Liu2011ApJ, Fu2011a, Comerford2012ApJ}, or source confused due to fiber spillover effects \citep{Husemann2018, Pfeifle2023ApJ} or extended narrow line regions \citep[e.g.,][]{H.Fu2012ApJ, H.Fu2018ApJ, Keel2019MNRAS}.
    
    The tidal subsamples reveal further insights. Those systems with large separations rarely exhibit tidal features, with only one case found at $>55\, h^{-1}_{70}$ kpc. As the projected distances decrease, the number of tidal systems rises markedly, peaking at 13 cases at $5-20\, h^{-1}_{70}$ kpc. 
    In the distribution of LOS velocity differences ($|\Delta v|$), the number of dual AGNs declines with increasing $|\Delta v|$, and nearly all systems showing tidal features are confined to $|\Delta v| < 300 \,\mathrm{km\,s^{-1}}$, consistent with the statistical trend reported by \citet{Liu2011ApJ}.
    These results collectively suggest that the tidal features become less common in wide separation systems, consistent with the predictions from the IllustrisTNG simulations, in which close pairs ($r<25$ kpc) have almost universally experienced recent pericentric passages and are expected to merge on short timescales ($\leq$1 Gyr; \citep{Patton2024MNRAS}), whereas wider pairs are more often long-lived or non-merging systems with weak or absent tidal signatures. 
    This indicates that larger physical distances may not favor the development or survival of tidal disturbances, possibly because (1) wide separation pairs are still in early merger stages where tidal features have not yet developed; (2) some systems have undergone close passages but with orbital or mass configurations not conducive to strong tidal signatures; or (3) galaxies have moved away from pericenter, and earlier tidal features have already dissipated. 
    It is also important to emphasize that the projected pair separation serves as a proxy for the merger stage, but in general it is a degenerate property; different merger stages/orbital configurations can show similar or the same projected separation \citep[e.g.,][]{Barnes1996, Cox2008, Capelo2015, Pfeifle2025}. In addition, the number of dual AGNs drops when $|\Delta v| \gtrsim 300\ \mathrm{km\ s^{-1}}$, consistent with the trend seen in \citet{Liu2011ApJ} sample, as mentioned in Section \ref{subsec31}.

    Beyond dual AGNs, we identified four triplets and one quadruplet system.
    These findings extend previous studies of multiple-AGN systems. Earlier works have reported a single bona fide, compact (sub-10 kpc) triple AGN system \citep{Pfeifle2019b, Liu2019ApJ3}, and confirmed Hickson Compact Group 16 as a triple AGN with larger separations \citep{Liu2011ApJ, Turner2001AA, Koss2012ApJL}.
    On galaxy-group scales, multiwavelength observations of compact groups \citep{DeRosa2015} have provided valuable insights into the mechanisms that trigger nuclear activity and star formation. Compared to these systems, our multiple-AGN samples exhibit separations of tens of kiloparsecs without pronounced tidal features, yet they represent a valuable intermediate-scale population for tracing the full evolutionary sequence of multiple-AGN activity and for probing the mechanisms that trigger nuclear accretion at different merger stages.

    Confirmed multiples remain exceedingly rare \citep{Turner2001AA, Koss2012ApJL, Liu2011ApJ3, Schawinski2011ApJ3, Liu2019ApJ3}. 
    Yet, their existence is also consistent with the hierarchical merging scenario, in which dense environments provide ample cold gas to fuel SMBH accretion \citep{Bhowmick2020MNRAS, Hennawi2006}. 
    Moreover, interactions among multiple SMBHs in these systems may accelerate SMBH growth and eventual coalescence \citep{Hoffman2023MNRAS}. 
    Therefore, multiple-AGN systems serve as ideal laboratories for understanding the co-evolution of SMBHs and their host galaxies \citep{Fukugita2004ApJ, Green2011ApJ, Husband2013MNRAS}.

    We also found some extra galaxies and/or AGN candidates by visual inspection. Detailed observations of these systems may reveal further cases of AGNs multiplets, as well as more information on the fueling and feedback processes of multi-AGNs in galaxy merger groups.   

\section{Conclusion} \label{sec5}
    In this work, we use the Million Quasar Catalog (MQC, version 8.0;\citealt{Flesch2023OJA})\footnote{\url{https://heasarc.gsfc.nasa.gov/W3Browse/all/milliquas.html}} as the parent catalog. From it, we construct a low-redshift ($z<0.5$) sample of dual AGNs and multiplets. Based on this sample, we perform statistical analyses. Our main conclusions are summarized as follows:
    
\begin{itemize}
    \item
    We applied stringent selection criteria, specifically (1) a projected separation of $r_p < 100\,h^{-1}_{70}$ kpc, and (2) a line-of-sight velocity difference of $|\Delta v| < 600\,\mathrm{km\,s^{-1}}$. Based on these criteria, we ultimately identified 173 dual AGNs, 4 AGN triplets, and 1 AGN quadruplet. Visual inspection of optical images from the DESI$-$LS were performed for each pair, revealing that approximately 16\% of the dual AGNs exhibit tidal features. 
    
    \item In our sample, AGN-NLAGNs (35.3\%) and NLAGN-NLAGNs (35.8\%) constitute the majority, indicating that narrow-line AGNs dominate in low-redshift dual AGN systems.
    The dual AGNs are generally uniformly distributed in projected distances between 0$-$100\( h^{-1}_{70} \, \text{kpc} \). The distribution of the tidal sample peaks sharply at \( r_p \approx 5-20 \, h^{-1}_{70} \, \text{kpc} \), which accounts for 13 pairs. The frequency then falls rapidly with increasing separation, with only a single pair located beyond 55 $\, h^{-1}_{70}$ kpc. Beyond 80$\, h^{-1}_{70}$ kpc, tidal pairs are entirely absent.
    
    \item 
    The redshift distribution of dual AGNs shows that within the range of $0.02 < z < 0.33$, the fraction of dual and multiple AGNs identified in the MQC is significantly lower than that reported by \citet{Liu2011ApJ} for their sample.

    \item 
    Regarding the LOS velocity difference ($|\Delta v|$), the number of dual AGNs decreases with increasing $|\Delta v|$, with only a few systems approaching the upper limit of 600 km s$^{-1}$. Systems exhibiting tidal features are predominantly found at $|\Delta v| < 300$ km s$^{-1}$, indicating that larger velocity differences generally correspond to greater physical separations, which are less conducive to the formation or persistence of prominent tidal structures.

    \item
    We discuss a close system, J1053+4710 ($r_p < 5 \,h^{-1}_{70}\,\mathrm{kpc}$), which shows a possible dual-nucleus morphology but the spectroscopic evidence is not sufficient to confirm a dual AGN nature.
    In addition, 165 wide separation pairs ($r_p > 10\,h^{-1}_{70}\,\mathrm{kpc}$) were identified, including 25 with clear tidal features, enriching the known population of large-scale interacting dual AGNs.

    \item
    We found some extra galaxies and/or AGN candidates in the same regions of some dual AGNs or multiplets, and some objects form interacting systems with the dual AGNs or multiplets. In total, we found 2 three-object, 3 four-object, 1 five-object, 1 six-object, 1 eight-object and 1 nine-object interacting systems.
\end{itemize}

    The MQC is inherently inhomogeneous, being compiled from diverse observational methods and archival datasets.
    Consequently, systematic biases and incompleteness may remain across different sky regions or redshift ranges, and the statistical results presented here should therefore be interpreted with caution.
    Nevertheless, with the advent of upcoming large-scale spectroscopic and photometric surveys such as DESI, 4MOST, and CSST, the number of confirmed dual and multiple AGNs is expected to increase substantially.
    These future surveys will enable the construction of a more complete and homogeneous sample, offering unprecedented opportunities to investigate the demographics, triggering mechanisms, and co-evolution of AGNs and their host galaxies across cosmic time.

\begin{acknowledgments}
We would like to express our heartfelt thanks to the anonymous referee for the careful and patient review and for the thoughtful, constructive comments that substantially improved the quality and clarity of the manuscript. 

This work has been supported by the Chinese National Natural Science Foundation grants 12333001, and by the National Key R\&D Program of China (2021YFA0718500, 2025YFA1614101). 

We would like to express our gratitude to Yue Han, for her unwavering support, and encouragement throughout the course of this research. Special thanks to Meicun Hou for her assistance during the research process.

\end{acknowledgments}

\newpage
\appendix
\setcounter{figure}{0}
\renewcommand{\thefigure}{\arabic{figure}}

\section*{Dual AGN sample mosaic} \label{appendix}
    Figure~\ref{fig:dual_mosaic} presents the mosaic of the dual AGN system identified in this work. 
    Each panel shows the $g$, $r$, and $z(i)$-band composite images from the DESI Legacy Surveys DR10, together with their corresponding model and residual images provided by the survey. These visualizations illustrate the morphological features and the model-fitting performance across our sample. It should be noted that some pairs lie at declinations lower than $-30^\circ$, where DESI imaging is not available, therefore, the displayed sample is not complete. 

    For the pairs J0114$-$5523 and J1419$+$5244, the MQC lists positions for two components, but only one component is apparent in the LS images at the expected locations. According to the MQC documentation\footnote{\url{http://www.quasars.org/Milliquas-ReadMe.txt}}, both objects are classified as type `N', a broad category that includes narrow-line AGNs, type-II Seyferts, and host-dominated systems, among other unquantified residue of legacy narrow-emission-line classes. This broad classification, together with potential astrometric uncertainties and/or a counterpart below the LS detection limit, may contribute to the apparent absence of the second component in the images.
    
    \begin{figure*}[htbp]
      \centering
      \includegraphics[width=\textwidth]{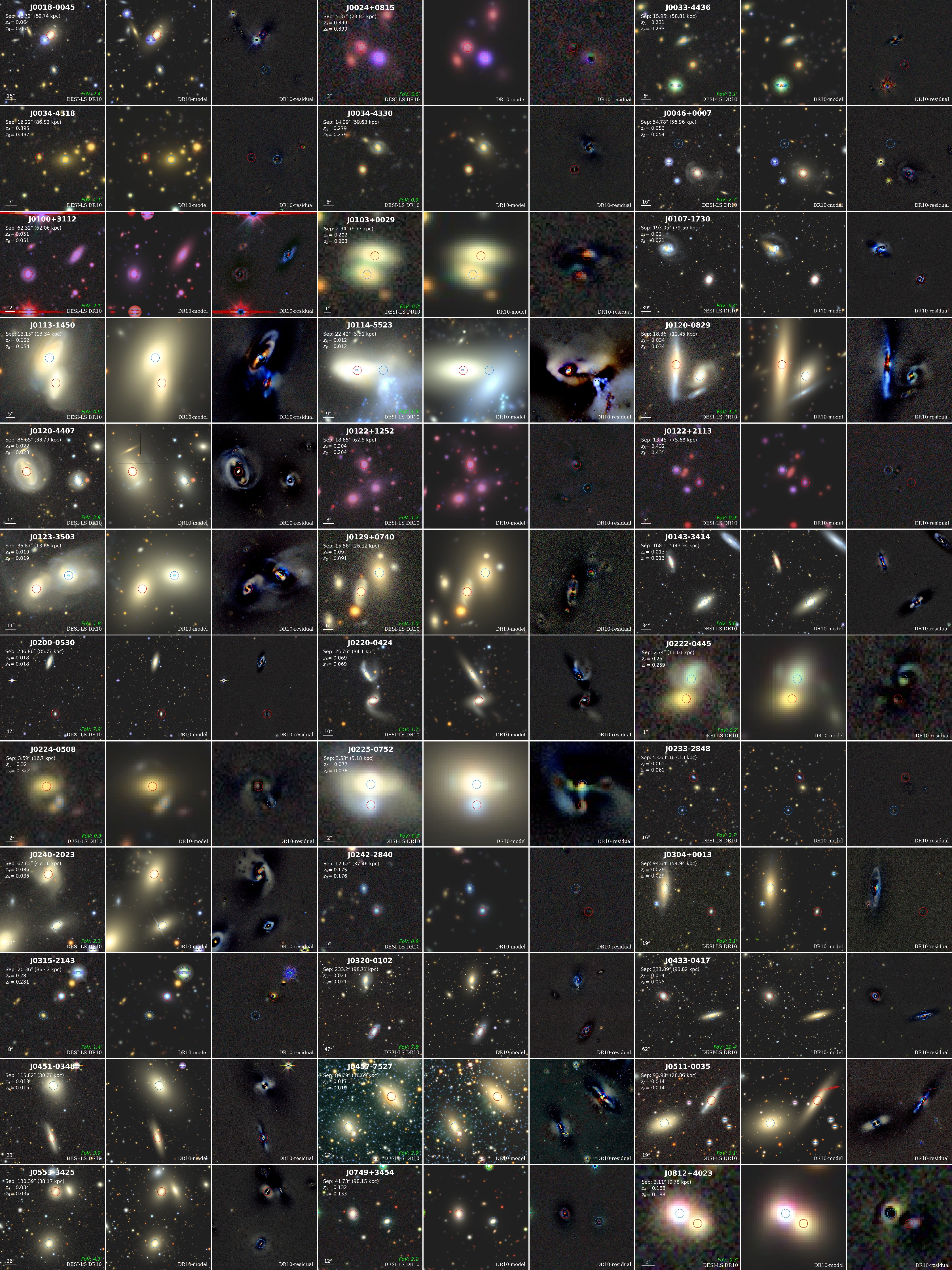}
      \caption{Mosaic of dual AGN systems. Each triplet of images shows the $grz(i)$-band composite, 
      the corresponding model, and the residual map from DESI Legacy Surveys DR10. The pinkish appearance in some images is likely due to incomplete band coverage. }
      \label{fig:dual_mosaic}
    \end{figure*}

    \begin{figure*}[htbp]
    \addtocounter{figure}{-1} 
    \centering
    \includegraphics[width=1\textwidth]{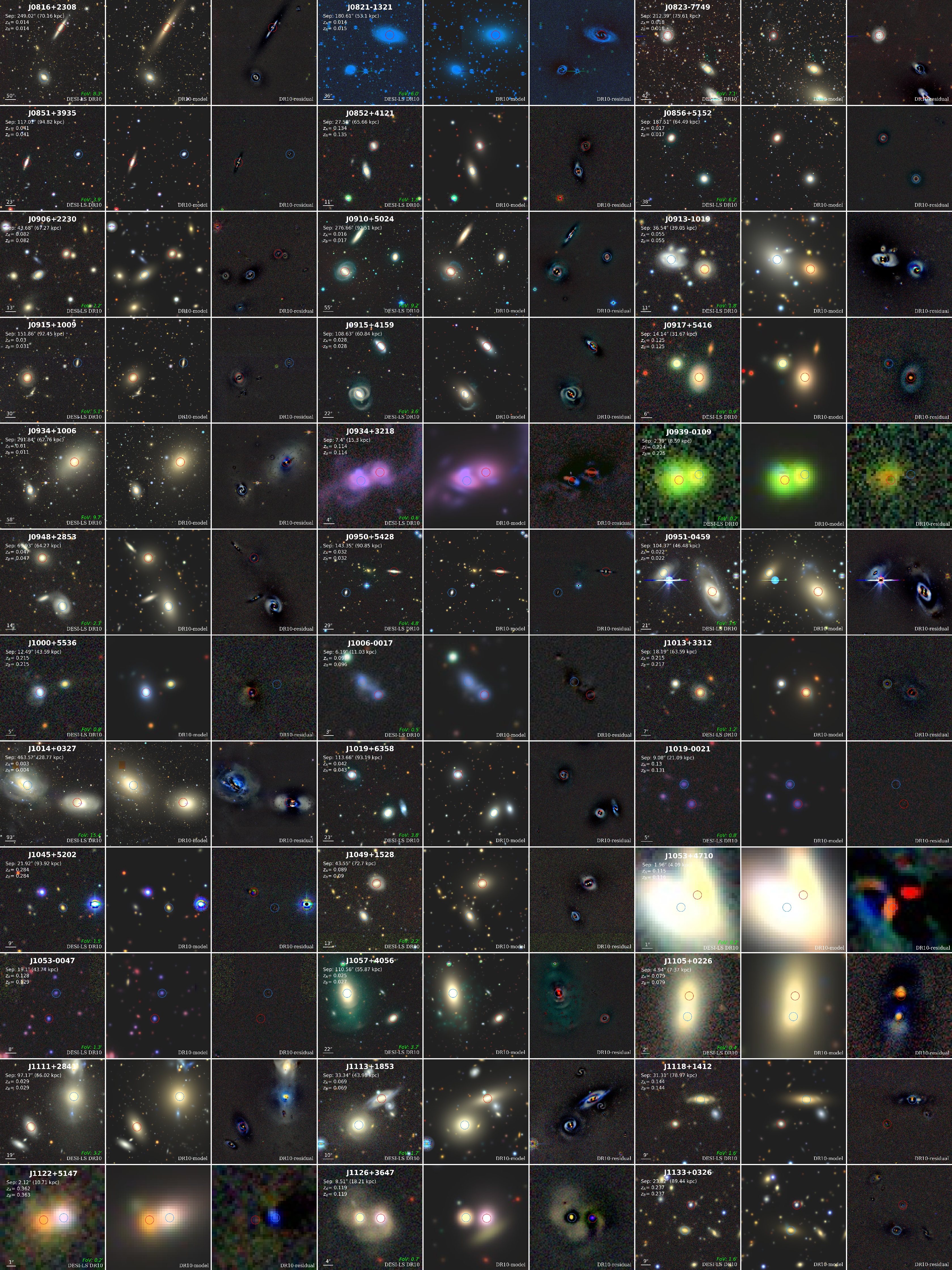}
    \caption{ - \textit{continued}}
    \end{figure*}

    \begin{figure*}[htbp]
    \addtocounter{figure}{-1} 
    \centering
    \includegraphics[width=1\textwidth]{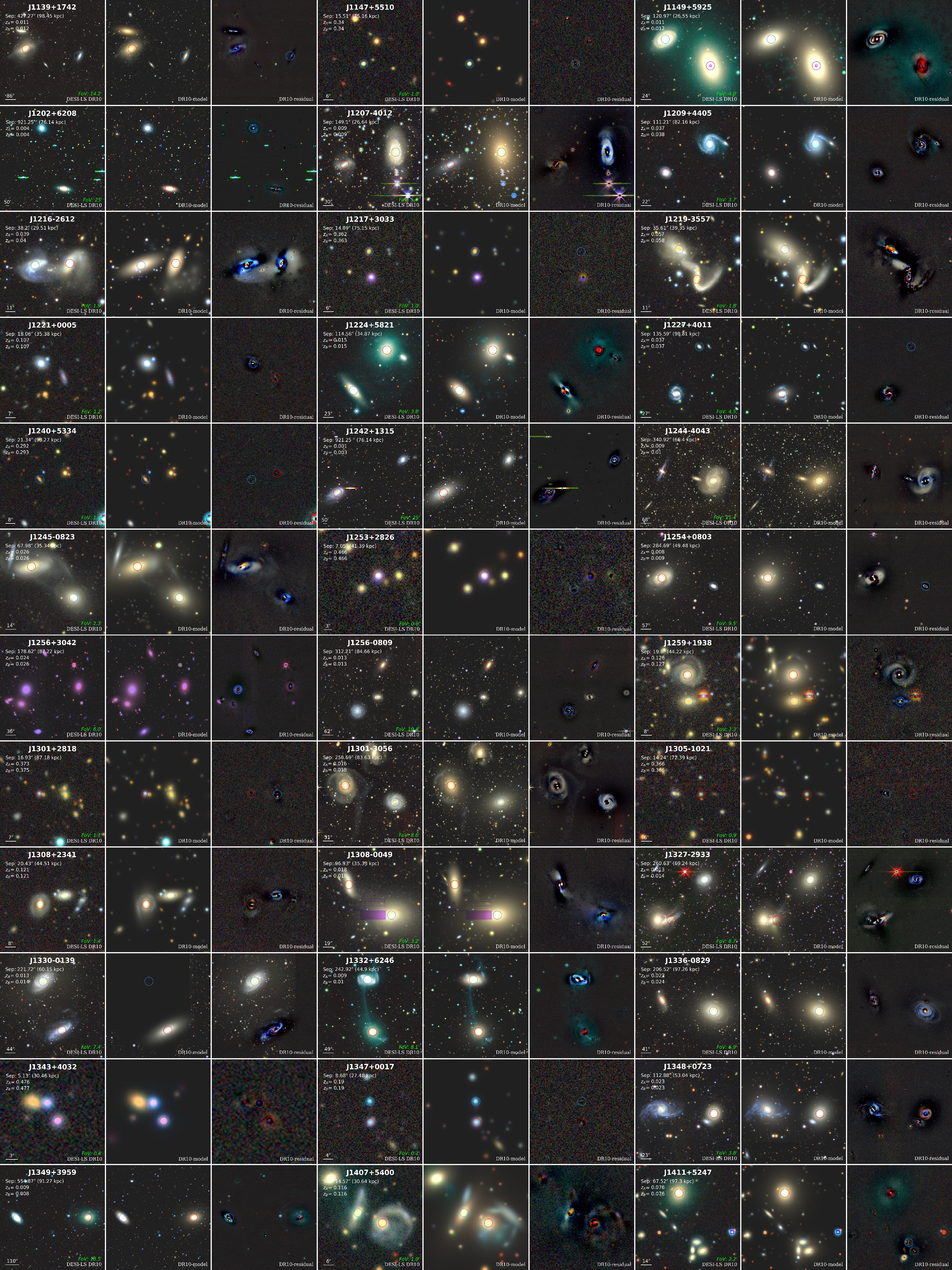}
    \caption{ - \textit{continued}}
    \end{figure*}

    \begin{figure*}[htbp]
    \addtocounter{figure}{-1} 
    \centering
    \includegraphics[width=1\textwidth]{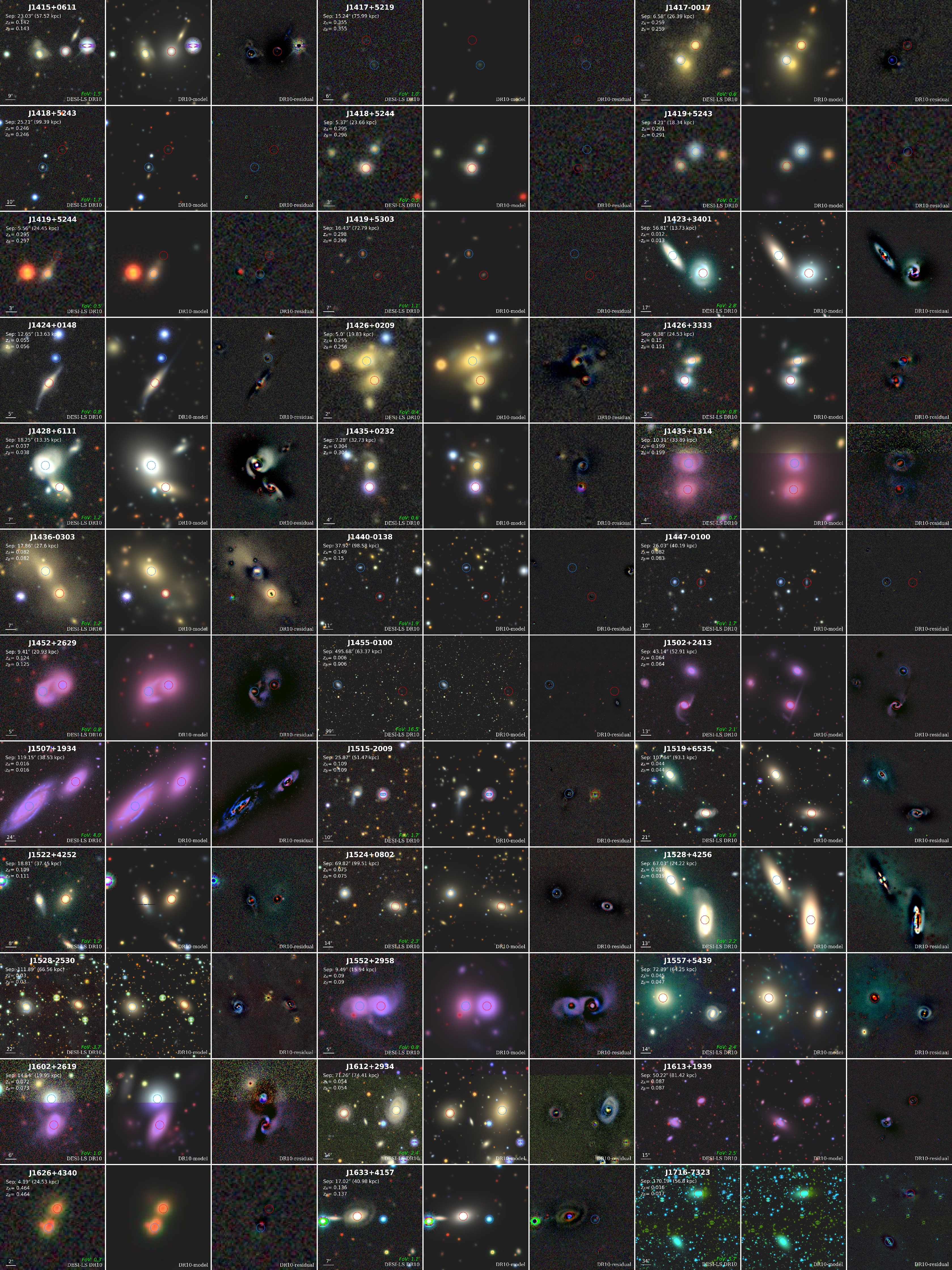}
    \caption{ - \textit{continued}}
    \end{figure*}

    \begin{figure*}[htbp]
    \addtocounter{figure}{-1} 
    \centering
    \includegraphics[width=1\textwidth]{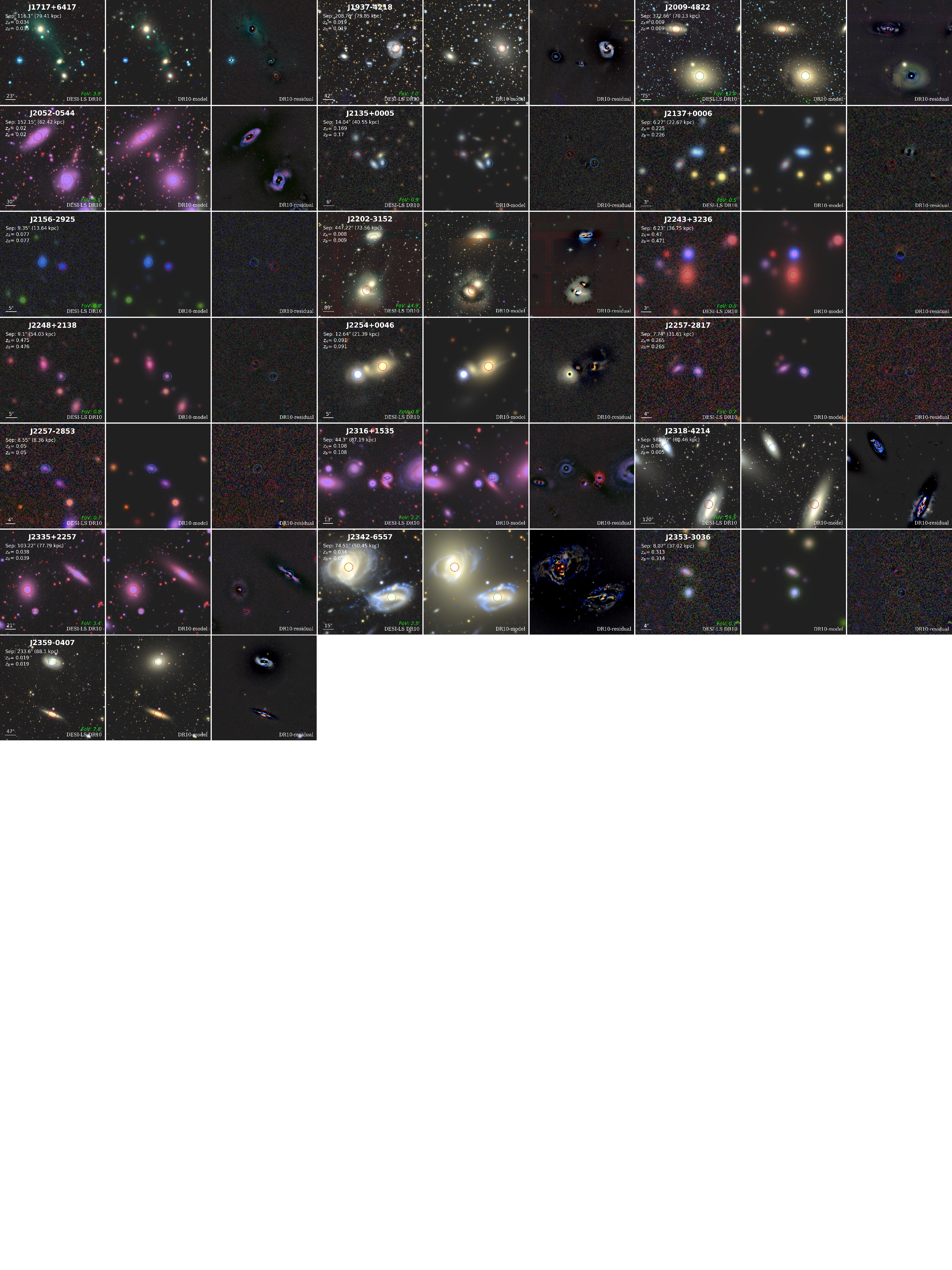}
    \caption{ - \textit{continued}}
    \end{figure*}

\newpage
\bibliography{sample701}{}
\bibliographystyle{aasjournalv7}

\end{document}